\documentclass[traditabstract]{aa}  
\usepackage{graphicx}
\usepackage{txfonts}
\usepackage{longtable,lscape}

\begin{document}

\title{Hot subdwarf stars in close-up view}

\subtitle{III. Metal abundances of subdwarf B stars \thanks{Table~A.1 is only available in electronic form at the CDS via anonymous ftp to cdsarc.u-strasbg.fr (130.79.128.5) or via http://cdsweb.u-strasbg.fr/cgi-bin/qcat?J/A+A/}}

\author{S. Geier \inst{1}}

\offprints{S.\,Geier,\\ \email{geier@sternwarte.uni-erlangen.de}}

\institute{Dr. Karl Remeis-Observatory \& ECAP, Astronomical Institute,
Friedrich-Alexander University Erlangen-Nuremberg, Sternwartstr. 7, D 96049 Bamberg, Germany}

\date{Received \ Accepted}

\abstract{ {\it Context} Hot subdwarf B stars (sdBs) are considered to be core helium-burning stars with very thin hydrogen envelopes situated on or near the extreme horizontal branch (EHB). The formation of sdBs is still unclear as well as the chemical composition of their atmospheres. The observed helium depletion is attributed to atmospheric diffusion. Metal abundances have been determined for about a dozen sdBs only resulting in puzzling patterns with enrichment of heavy metals and depletion of lighter ones. {\it Aims} In this paper we present a detailed metal abundance analysis of 106 sdBs. {\it Methods} From high resolution spectra we measured elemental abundances of up to 24 different ions per star. A semi-automatic analysis pipeline was developed to calculate and fit LTE models to a standard set of spectral lines. {\it Results} A general trend of enrichment was found with increasing temperature for most of the heavier elements. The lighter elements like carbon, oxygen and nitrogen are depleted and less affected by temperature. Although there is considerable scatter from star to star, the general abundance patterns in most sdBs are similar. State-of-the-art diffusion models predict such patterns and are in qualitative agreement with our results. However, the highest enrichments measured cannot not be explained with these models. Peculiar line shapes of the strongest metal lines in some stars indicate vertical stratification to be present in the atmospheres. Such effects are not accounted for in current diffusion models and may be responsible for some of the yet unexplained abundance anomalies.
}

\maketitle

\section{Introduction \label{sec:intro}}

Subluminous B (sdB) stars are considered to be core helium-burning stars with very thin hydrogen envelopes and masses around $0.5\,M_{\rm \odot}$ (Heber et al. \cite{heber86}; Heber \cite{heber09}). The formation of these objects is still unclear. SdB stars can only be formed, if the progenitor loses its envelope almost entirely after passing the red-giant stage and the remaining hydrogen-rich envelope has not retained enough mass to sustain a hydrogen-burning shell. The star cannot ascend the asymptotic giant branch, but remains on the extreme horizontal branch (EHB) until the core helium-burning stops and eventually evolves to become a white dwarf. The reason for this very high mass loss near the core helium flash is still unclear. Several single star and close binary scenarios are currently under discussion (see Han et al. \cite{han02}, \cite{han03}; Yi \cite{yi08} and references therein). Close binary evolution is a promising option, because the envelope can be lost through common envelope ejection or stable Roche lobe overflow. An alternative way of forming a single sdB is the merger of two helium white dwarfs (Webbink \cite{webbink84}; Iben \& Tutukov \cite{iben84}). 

\begin{figure}[t!]
	\resizebox{\hsize}{!}{\includegraphics{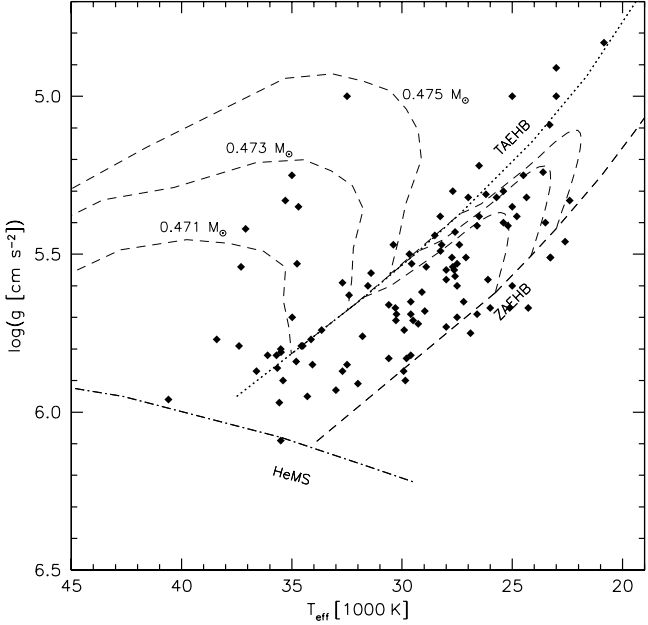}}
	\caption{Surface gravity is plotted against effective temperature. The diamonds mark the programme stars from our sample. Zero Age Extreme Horizontal Branch (ZAEHB) and Terminal Age Extreme Horizontal Branch (TAEHB) are plotted as well as the Helium Main Sequence (HeMS). Tracks for canonical EHB stars are taken from Dorman et al. (\cite{dorman93}). The sample covers the whole parameter range on the EHB and also subdwarfs that already evolved away from the EHB.}
	\label{teffloggsdB}
\end{figure}

The abundance anomalies of hot subdwarf stars are very important to understand the physical properties of hot stellar atmospheres. Sargent \& Searle (\cite{sargent66}) discovered the helium deficiency of sdB stars for the first time. This finding seemed to be at variance with the big-bang theory of nucleosynthesis. Greenstein, Truran \& Cameron (\cite{greenstein74}) suggested that diffusion in the hot atmosphere of sdBs could cause the helium deficiency. Peculiar metal abundances were first reported by Baschek et al. (\cite{baschek72}) for HD\,4539. While some metals show  solar abundances, others are depleted or even enriched. 

Studies of optical spectra remained scarce. Heber et al. (\cite{heber99}, \cite{heber00}) analysed high resolution spectra of four pulsating sdB stars taken with Keck/HIRES. Napiwotzki et al. (\cite{napiwotzki01}) and Telting et al. (\cite{telting08}) derived metal abundances of the sdB binary HE\,1047$-$0436 and the hybrid sdB pulsator Baloon\,090100001 from optical spectra, respectively. Finally, Edelmann et al. (\cite{edelmann99,edelmann01}), Przybilla et al. (\cite{przybilla06}) as well as Pereira \& Jeffery (\cite{pereira08}) published some preliminary results of their analysis of high resolution spectra. 

Hot stars display a much larger number of spectral lines in the UV than in the optical.
With the advent of the IUE satellite it became possible to determine abundances of C, N and Si from strong UV resonance lines for nine sdB stars (Lamontagne et al. \cite{lamontagne87} and references therein). Lines of heavier metal have been analysed from IUE spectra of two stars (Baschek et al. \cite{baschek82a}, \cite{baschek82b}). The most detailed analysis of UV spectra has been carried out by O'Toole \& Heber (\cite{otoole06}) based on spectra obtained with STIS onboard HST. The FUSE satellite opened up the FUV-regime for abundance studies. Many sdBs have been observed, but only some have been analysed (Fontaine et al. \cite{fontaine06}; Chayer et al. \cite{chayer06}; Blanchette et al. \cite{blanchette08}; Pereira \& Jeffery \cite{pereira08}). 

Due to their faintness, quantitative analyses of hot subdwarfs in globular clusters (GCs) are restricted to low-resolution spectroscopy. Therefore constraints can only be put on the helium abundances and the analysis of metal abundances remains coarse (Heber et al. \cite{heber86b}; Moni Bidin et al. \cite{monibidin07}; Moehler et al. \cite{moehler07}; Brown et al. \cite{brown11}).

The Blue Horizontal Branch (BHB) stars are the cooler siblings of the EHB stars ($T_{\rm eff}<20\,000\,{\rm K}$; for a review see Behr \cite{behr03a}). Their chemical composition is of interest especially in GC research. Since the morphology of the HB in GCs still remains unclear, different explanations for the shape of and the apparent gaps along the HB have been proposed. In GCs all stars belong to the same population and should therefore have similar primordial chemical compositions. 

Glaspey et al. (\cite{glaspey89}) were the first to discover a significant change of chemical abundances as function of the position on the HB. While a BHB star with $T_{\rm eff}\simeq10\,000\,{\rm K}$ in NGC\,6752 showed helium and iron abundances similar to the cluster composition (which is usually derived from abundance studies of red giants), a hotter one ($T_{\rm eff}\simeq16\,000\,{\rm K}$) turned out to show depletion of helium and strong enrichment of iron. Further abundance studies of BHB stars in several GCs revealed a general pattern (Moehler et al. \cite{moehler99}; Behr et al. \cite{behr03a}; Fabbian et al. \cite{fabbian05}; Pace et al. \cite{pace06}), which can also be observed in field BHB stars (Behr et al. \cite{behr03b}; For \& Sneden \cite{for10}). Stars cooler than about $11\,500\,{\rm K}$ show the typical abundances of their parent population, while stars hotter than that are in general depleted in helium and strongly enriched in iron and other heavy elements like titanium or chromium. Lighter elements like magnesium and silicon on the other hand have normal abundances.

Diffusion processes in the stellar atmosphere are responsible for this effect. Michaud et al. (\cite{michaud89}) predicted such abundance patterns even before the anomalies were observed. Caloi (\cite{caloi99}) explained the sharp transition between the two abundance patterns as disappearance of subsurface convection layers at a critical temperature. Modeling BHB stars Sweigart (\cite{sweigart97b}) indeed found that thin convective layers below the surface driven by hydrogen ionisation should exist and move closer to the surface, as soon as the temperature increases. At about $12\,000\,{\rm K}$ the convection zone reaches the surface and the outer layer of the star becomes fully radiative. Since convection is very efficient in mixing the envelope, diffusion processes cannot set in below this limit. In hotter stars with radiative atmospheres helium is expected to diffuse downward, since its mean molecular weight is higher than the one of hydrogen. Heavier elements on the other hand present sufficiently large cross sections to the outgoing radiation field and experience radiative accelerations greater than gravity. Hence these elements become enriched in the atmosphere. If the radiative acceleration almost equals gravity, the diffusion timescales get very long and the element is not significantly affected by diffusion. Michaud et al. (\cite{michaud08}) modelled these effects and reproduced for the first time the observed abundance patterns of BHB stars.

Atmospheric diffusion processes have also been invoked to explain abundance pecularities in a wide range of stars including white dwarfs, luminous blue variables, low mass halo stars, Ap and Am stars, and HgMn stars (see Vauclair \& Vauclair \cite{vauclair82} for a review). For sdB stars the first theoretical diffusion models met with little success only (e.g. Michaud et al. \cite{michaud83}). Since then several attempts have been made to model the atmospheres of sdBs (Bergeron et al. \cite{bergeron88}; Michaud et al. \cite{michaud89}; Fontaine \& Chayer \cite{fontaine97}; Ohl et al. \cite{ohl00}; Unglaub \& Bues \cite{unglaub01}). Radiative levitation and mass loss caused by stellar winds (Vink \& Cassisi \cite{vink02}) have been invoked to counteract the gravitational settling as well as extra mixing at the surface (Michaud et al. \cite{michaud11}; Hu et al. \cite{hu11}). 

Here we present a metal abundance analysis of 106 sdB stars, by far the largest sample to date. Previous papers of the this series dealt with the rotational properties of sdB binaries (Geier et al. \cite{geier10}, Paper I) and the rotational properties of single sdBs (Geier \& Heber \cite{geier12a}, Paper II). 

In Sect. \ref{sec:obs} the dataset is described. Sect.~\ref{sec:ana} introduces the semi-automatic analysis pipeline used to measure elemental abundances. In Sect.~\ref{sec:lit} our results are compared the with literature and systematic uncertainties are discussed. The metal abundances, general trends as well as details for individual elements, are presented in Sect.~\ref{sec:metal}. The abundance patterns of different sdB sub-populations (e.g pulsators vs. non-pulsators) are discussed in Sects.~\ref{sec:sub}, \ref{sec:helium} and \ref{sec:other}, while Sect.~\ref{sec:bhb} contains a comparison of our results with the abundance patterns on the blue horizontal branch. State-of-the-art diffusion models are compared to our results in Sect.~\ref{sec:lev}. Remaining abundance anomalies and peculiar shapes of metal lines are discussed in (Sect.~\ref{sec:strat}). Finally, conclusions are drawn in Sect.~\ref{sec:con}.

\begin{figure}[t!]
	\resizebox{\hsize}{!}{\includegraphics{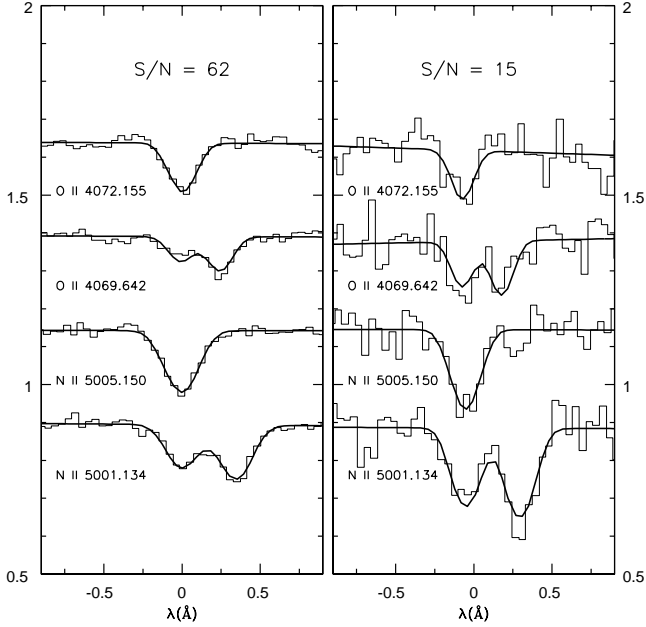}}
	\caption{Example fits of common oxygen and nitrogen lines for two spectra with different quality. Although the fit quality deteriorates, it is still possible to fit models with reasonable equivalent widths.}
	\label{quality}
\end{figure}

\section{Sample of sdBs with high resolution spectra \label{sec:obs}}

We selected a large sample of 106 sdB stars for which high resolution spectroscopy ($R=20\,000-48\,000$) suitable for the analysis of weak and sharp metal lines is available. Observations and data reduction are described in Lisker et al. (\cite{lisker05}), Geier et al. (\cite{geier12b}) and the Papers I and II of this series. 

Our sample contains $38$ radial velocity (RV) variable systems in close binary systems. The remaining $68$ are apparently single stars. Eleven sdBs in our sample are known pulsators. Four of them belong to the class of short-period pulsators (V361\,Hya, sdBV$_{\rm r}$), seven are long-period pulsators (V1093\,Her, sdBV$_{\rm s}$). Our programme stars cover the entire parameter range in the $T_{\rm eff}-\log{g}$-diagram (see Fig.~\ref{teffloggsdB}).

\begin{table}[t!]
\caption{Standard line list}  \label{tab:lines}
\begin{tabular}{llllll}
\hline
\noalign{\smallskip}
Ion              & $\lambda\,{\rm [\AA]}$   &  Ion              & $\lambda\,{\rm [\AA]}$   & Ion              & $\lambda\,{\rm [\AA]}$   \\
\noalign{\smallskip}                           
\hline                                                            
\noalign{\smallskip}                                              
Ne\,{\sc ii}  &     3694.212  & V\,{\sc iii}    &    4274.417  &  S\,{\sc ii}     &    4815.552 \\
Ne\,{\sc ii}  &     3713.080  & Ti\,{\sc iii}   &    4281.555  &  Si\,{\sc iii}   &    4828.951 \\
Si\,{\sc iii} &     3796.124  & S\,{\sc iii}    &    4284.979  &                  &    4829.030 \\ 
              &     3796.203  & Fe\,{\sc iii}   &    4286.091  &                  &    4829.111 \\ 
Si\,{\sc iii} &     3806.562  &                 &    4286.128  &  Ar\,{\sc ii}  &     4879.863 \\  
              &     3806.700  &                 &    4286.164  &  O\,{\sc ii}   &     4890.858 \\  
O\,{\sc ii}   &     3850.799  & V\,{\sc iii}    &    4294.919  &  O\,{\sc ii}   &     4906.833 \\
              &     3851.033  &  O\,{\sc ii}    &    4303.615  &  O\,{\sc ii}   &     4941.069 \\
O\,{\sc ii}   &     3911.959  &                 &    4303.833  &  O\,{\sc ii}   &     4943.003 \\
              &     3912.107  & Fe\,{\sc iii}   &    4304.748  &  N\,{\sc ii}   &     4994.353 \\
C\,{\sc ii}   &     3920.681  &                 &    4304.767  &                &     4994.360 \\
O\,{\sc ii}   &     3945.038  &  O\,{\sc ii}    &    4342.009  &                &     4994.370 \\
N\,{\sc ii}   &     3994.997  &  O\,{\sc ii}    &    4349.426  &  N\,{\sc ii}   &     5001.134 \\
N\,{\sc ii}   &     4035.081  &  O\,{\sc ii}    &    4351.262  &                &     5001.474 \\
N\,{\sc ii}   &     4041.310  & Fe\,{\sc iii}   &    4352.577  &  N\,{\sc ii}   &     5005.150 \\
N\,{\sc ii}   &     4043.532  & Fe\,{\sc iii}   &    4371.337  &  N\,{\sc ii}   &     5007.328 \\
C\,{\sc iii}  &     4056.061  & V\,{\sc iii}    &    4383.391  &  Sc\,{\sc iii} &     5032.072 \\
P\,{\sc iii}  &     4059.312  & Ne\,{\sc ii}    &    4391.991  &  Fe\,{\sc iii} &     5063.421 \\
O\,{\sc ii}   &     4060.526  &  O\,{\sc ii}    &    4414.905  &  Si\,{\sc iii} &     5091.250 \\
Sc\,{\sc iii} &     4061.210  &  O\,{\sc ii}    &    4416.974  &                &     5091.364 \\
O\,{\sc ii}   &     4069.623  & Fe\,{\sc iii}   &    4419.596  &                &     5091.455 \\
              &     4069.886  & Mg\,{\sc ii}    &    4481.126  &                &     5091.543 \\
O\,{\sc ii}   &     4072.157  &                 &    4481.325  &  Fe\,{\sc iii} &     5127.387 \\
P\,{\sc iii}  &     4080.089  &  O\,{\sc ii}    &    4452.375  &                &     5127.631 \\
N\,{\sc ii}   &     4082.270  & Al\,{\sc iii}   &    4528.945  &  C\,{\sc ii}   &     5132.947 \\
Si\,{\sc iv}  &     4088.862  &                 &    4529.189  &                &     5133.281 \\
O\,{\sc ii}   &     4097.258  &  N\,{\sc ii}    &    4530.410  &  C\,{\sc ii}   &     5143.495 \\
Si\,{\sc iv}  &     4116.104  & Si\,{\sc iii}   &    4552.622  &  C\,{\sc ii}   &     5145.165 \\
O\,{\sc ii}   &     4119.215  & Si\,{\sc iii}   &    4567.840  &  C\,{\sc ii}   &     5151.085 \\
Fe\,{\sc iii} &     4122.780  & Si\,{\sc iii}   &    4574.757  &  Fe\,{\sc iii} &     5156.111 \\
K\,{\sc ii}   &     4134.723  &  O\,{\sc ii}    &    4590.972  &  N\,{\sc ii}   &     5175.896 \\
Fe\,{\sc iii} &     4137.764  &  O\,{\sc ii}    &    4596.175  &  N\,{\sc ii}   &     5179.521 \\
Fe\,{\sc iii} &     4139.350  &  C\,{\sc ii}    &    4618.559  &  Fe\,{\sc iii} &     5193.909 \\
Al\,{\sc iii} &     4149.913  &                 &    4619.249  &  S\,{\sc ii}   &     5212.267 \\
              &     4149.968  &  N\,{\sc ii}    &    4630.539  &                &     5212.620 \\
              &     4150.173  &  N\,{\sc iii}   &    4634.126  &  Fe\,{\sc iii} &     5235.658 \\
C\,{\sc iii}  &     4162.877  &  N\,{\sc iii}   &    4640.644  &  Fe\,{\sc iii} &     5276.476 \\
Fe\,{\sc iii} &     4164.731  &  O\,{\sc ii}    &    4641.810  &  Fe\,{\sc iii} &     5282.297 \\
              &     4164.916  &  C\,{\sc iii}   &    4647.418  &                &     5282.579 \\
Fe\,{\sc iii} &     4166.840  &  O\,{\sc ii}    &    4649.134  &  Fe\,{\sc iii} &     5299.926 \\
N\,{\sc ii}   &     4171.595  &  C\,{\sc iii}   &    4650.246  &  Fe\,{\sc iii} &     5302.602 \\
S\,{\sc ii}   &     4174.265  &                 &    4651.016  &  S\,{\sc ii}   &     5320.723 \\
N\,{\sc ii}   &     4176.195  &                 &    4651.473  &  C\,{\sc ii}   &     5342.376 \\
K\,{\sc ii}   &     4186.162  &  Si\,{\sc iv}   &    4654.312  &  S\,{\sc ii}   &     5345.712 \\
O\,{\sc ii}   &     4189.789  &  O\,{\sc ii}    &    4661.633  &                &     5346.084 \\
N\,{\sc iii}  &     4195.760  &  O\,{\sc ii}    &    4676.235  &  Ti\,{\sc iv}  &     5398.930 \\
Ti\,{\sc iii} &     4207.491  &  N\,{\sc ii}    &    4678.135  &  S\,{\sc ii}   &     5432.797 \\
Ti\,{\sc iii} &     4215.525  &  N\,{\sc ii}    &    4694.642  &  Ti\,{\sc iv}  &     5492.512 \\
Fe\,{\sc iii} &     4222.271  &  O\,{\sc ii}    &    4699.003  &  N\,{\sc ii}   &     5535.346 \\
Ca\,{\sc iii} &     4233.713  &                 &    4699.220  &  S\,{\sc ii}   &     5606.151 \\
              &     4233.736  &  O\,{\sc ii}    &    4701.184  &  N\,{\sc ii}   &     5679.558 \\
N\,{\sc ii}   &     4236.927  &                 &    4701.708  &  N\,{\sc ii}   &     5686.213 \\
              &     4237.047  &  O\,{\sc ii}    &    4703.163  &  N\,{\sc ii}   &     5710.766 \\
Ca\,{\sc iii} &     4240.742  &  O\,{\sc ii}    &    4705.352  &  Si\,{\sc iii} &     5739.734 \\
N\,{\sc ii}   &     4241.755  & Si\,{\sc iii}   &    4716.654  &  Fe\,{\sc iii} &     5833.938 \\
              &     4241.786  & Ar\,{\sc ii}    &    4735.906  &  N\,{\sc ii}   &     5893.147 \\
S\,{\sc iii}  &     4253.589  &  C\,{\sc ii}    &    4737.966  &  N\,{\sc ii}   &     5931.782 \\
K\,{\sc ii}   &     4263.447  &  C\,{\sc ii}    &    4744.766  &  N\,{\sc ii}   &     5941.654 \\
C\,{\sc ii}   &     4267.001  &  C\,{\sc ii}    &    4747.279  &  Fe\,{\sc iii} &     6032.604 \\
              &     4267.261  & Ar\,{\sc ii}    &    4806.021  &  N\,{\sc ii}   &     6167.755 \\
V\,{\sc iii}  &     4268.183  & Si\,{\sc iii}   &    4813.333  &           &              \\
\noalign{\smallskip}                           
\hline                                         
\end{tabular}
\end{table}

\section{Semi-automatic abundance analysis \label{sec:ana}}

In order to derive the metal abundances we compared the observed spectra with synthetic line profiles. Metal line-blanketed LTE (local thermodynamic equilibrium) model atmospheres (Heber et al. \cite{heber00}) were computed for the atmospheric parameters given in Lisker et al. (\cite{lisker05}), Geier et al. (\cite{geier12b}) and Paper I and II using the LINFOR program (developed by Holweger, Steffen and Steenbock at Kiel university, modified by Lemke \cite{lemke97}).

A standard set of lines was chosen taking into account several criteria. First of all the lines had to be strong enough to be detectable in noisy spectra. Blends with lines of different ions were not used. Only line blends of the same ion could be handled because only one abundance was fitted to individual lines or multiplets. We selected a set of 182 metal lines from 24 different ions (see Table \ref{tab:lines}) and used atomic data from the lists of Kurucz (\cite{kurucz92}), Wiese et al. (\cite{wiese96}), Ekberg (\cite{ekberg93}), and Hirata \& Horaguchi (\cite{hirata95}). For carbon, nitrogen, oxygen and silicon, the NIST database was used to obtain state-of-the-art atomic data. 

\begin{figure}[t!]
	\resizebox{\hsize}{!}{\includegraphics{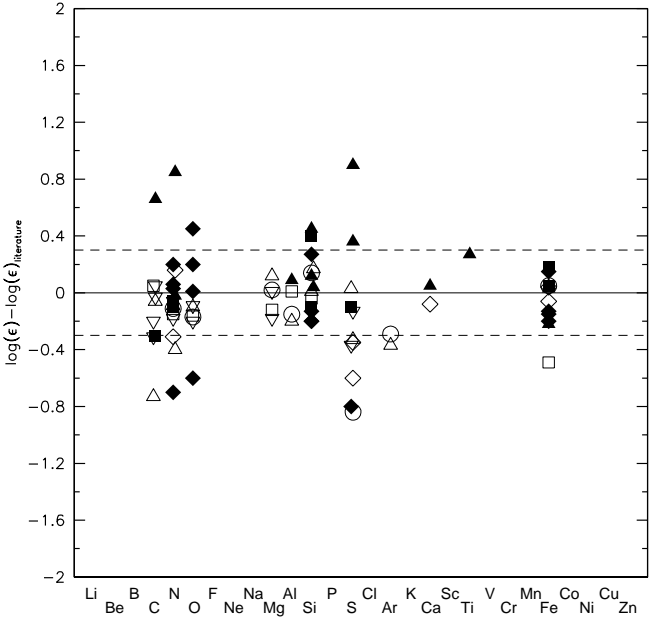}}
	\caption{The residuals between the metal abundances determined here and from literature are plotted against the chemical elements. Filled symbols mark results derived from UV-spectroscopy (squares, Fontaine et al. \cite{fontaine06}; diamonds, Blanchette et al. \cite{blanchette08}; triangles, O'Toole \& Heber \cite{otoole06}), open symbols results derived from optical high-resolution spectra (circles, Edelmann et al. \cite{edelmann99}; upward triangles, Napiwotzki et al. \cite{napiwotzki01}; squares, Pereira \& Jeffery \cite{pereira08}; diamonds, Heber et al. \cite{heber00}; downward triangles, Przybilla et al. \cite{przybilla06}). The dashed horizontal lines mark our statistical uncertainty estimate ($\pm0.3\,{\rm dex}$).}
	\label{abun_residuals}
\end{figure}

A simultaneous fit of elemental abundance, projected rotational velocity ($v_{\rm rot}\sin{i}$) and radial velocity (RV) was then performed for each identified line using the FITSB2 routine (Napiwotzki et al. \cite{napiwotzki04}). Inappropiate lines were neglected. This rejection procedure included several criteria. Equivalent width and depth of the fitted line was measured and compared to the noise level to distinguish between lines and noise features. The resulting individual radial velocity had to be low, because all spectra were corrected to zero RV before. Otherwise the lines were considered as misidentifications or noise features. Then the fit quality given by the $\chi^2$ had to be comparable to the average value to sort out lines contaminated by blends or artifacts caused by cosmic rays. Mean value and standard deviation were calculated from all abundance measurements of each ion. Because not all lines were present or suitable for fits in each star, the number of fitted lines differs. Upper limits were calculated by comparing the depth of the rotationally broadened synthetic spectral lines with the noise level. If only one line was found suitable for determining the abundance, the upper limits derived for the other lines of the same element where compared to this abundance. Only consistent results (within the error margins) were considered reliable, lowering the probality of misidentifications.

The accuracy of our results is limited by the quality of the spectra. Fig. \ref{quality} shows two examples of spectra with highly different quality. The errors given in Table~A.1 are the standard deviations of the individual line measurements. Numerical experiments were carried out to quantify the impact of noise on the result (see Paper I). We therefore regard $0.3\,{\rm dex}$ as typical statistical uncertainty of our abundance analysis. Some lines have peculiar profiles (see Sect.~\ref{sec:strat}), which cannot be matched with the synthetic models. However, the equivalent widths are similar to the models and the abundance determination should therefore be correct to within $\pm0.5\,{\rm dex}$.

\begin{figure*}[t!]
\begin{center}
	\resizebox{13cm}{!}{\includegraphics{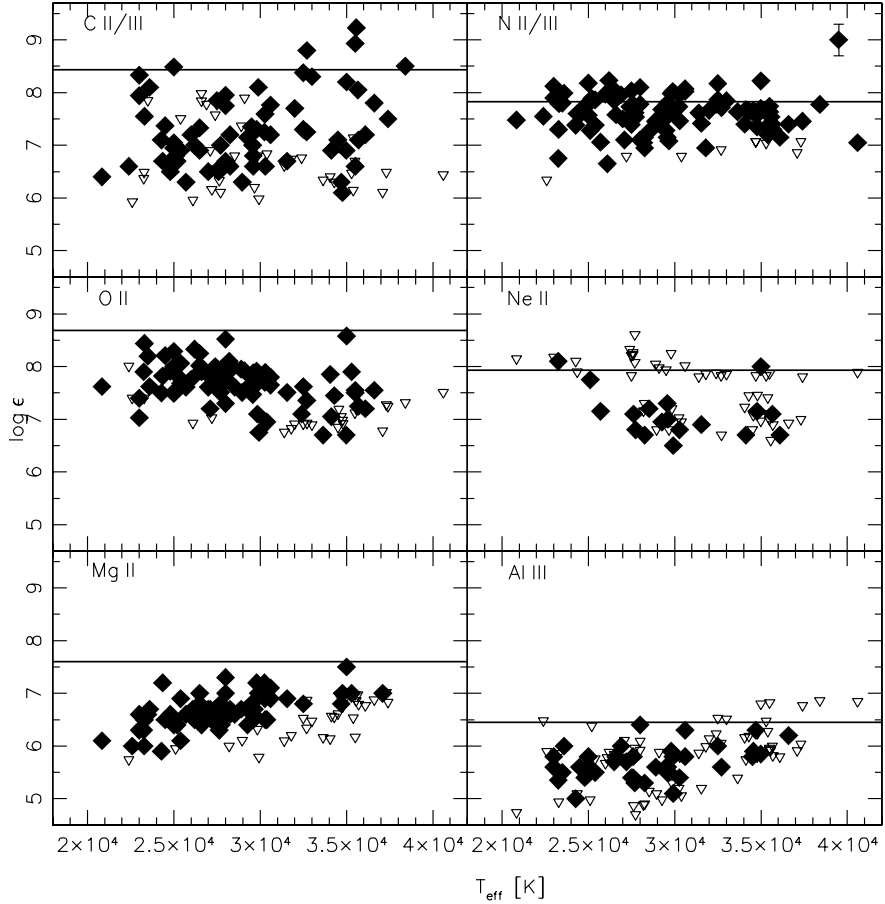}}
	\caption{Elemental abundances from carbon to aluminium plotted against effective temperature. If two ionisation stages are present, the average abundance is given. The filled diamonds mark measured abundances while the open triangles mark upper limits. Typical error bars are given in the upper right corner. The solid horizontal lines mark solar abundances (Asplund et al. \cite{asplund09}).}
	\label{abuns1}
\end{center}
\end{figure*}

\begin{figure*}[t!]
\begin{center}
	\resizebox{13cm}{!}{\includegraphics{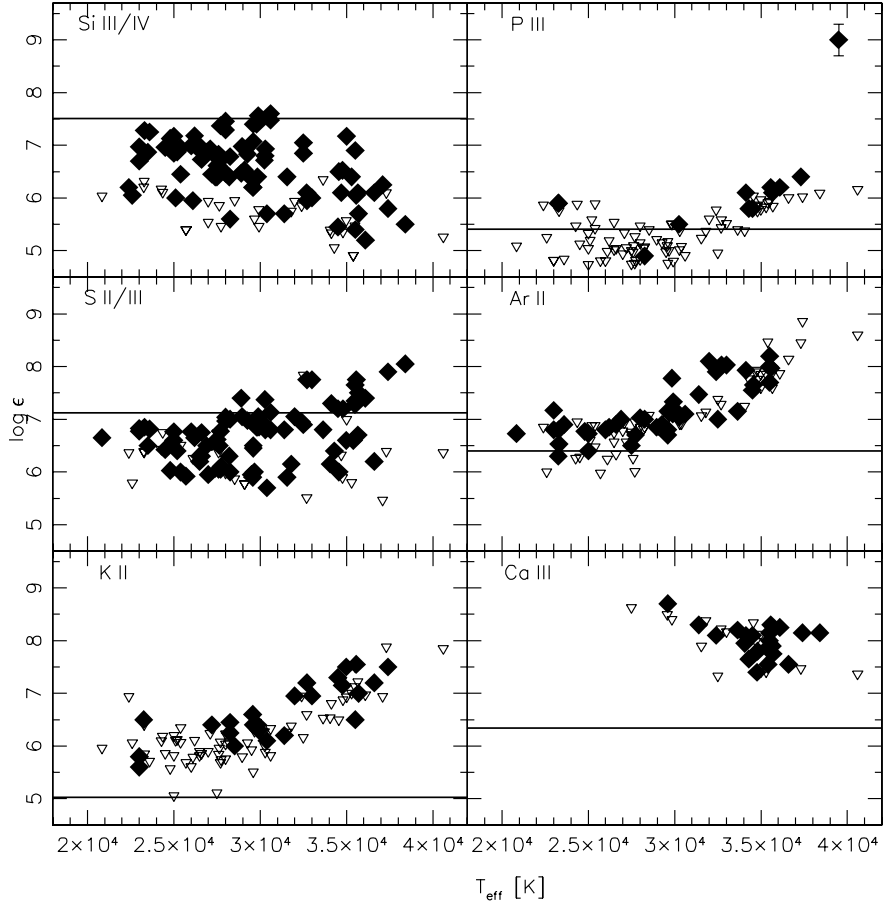}}
	\caption{Elemental abundances from silicon to calcium plotted against effective temperature (see Fig.~\ref{abuns1}).}
	\label{abuns2}
\end{center}
\end{figure*}

\begin{figure*}[t!]
\begin{center}
	\resizebox{13cm}{!}{\includegraphics{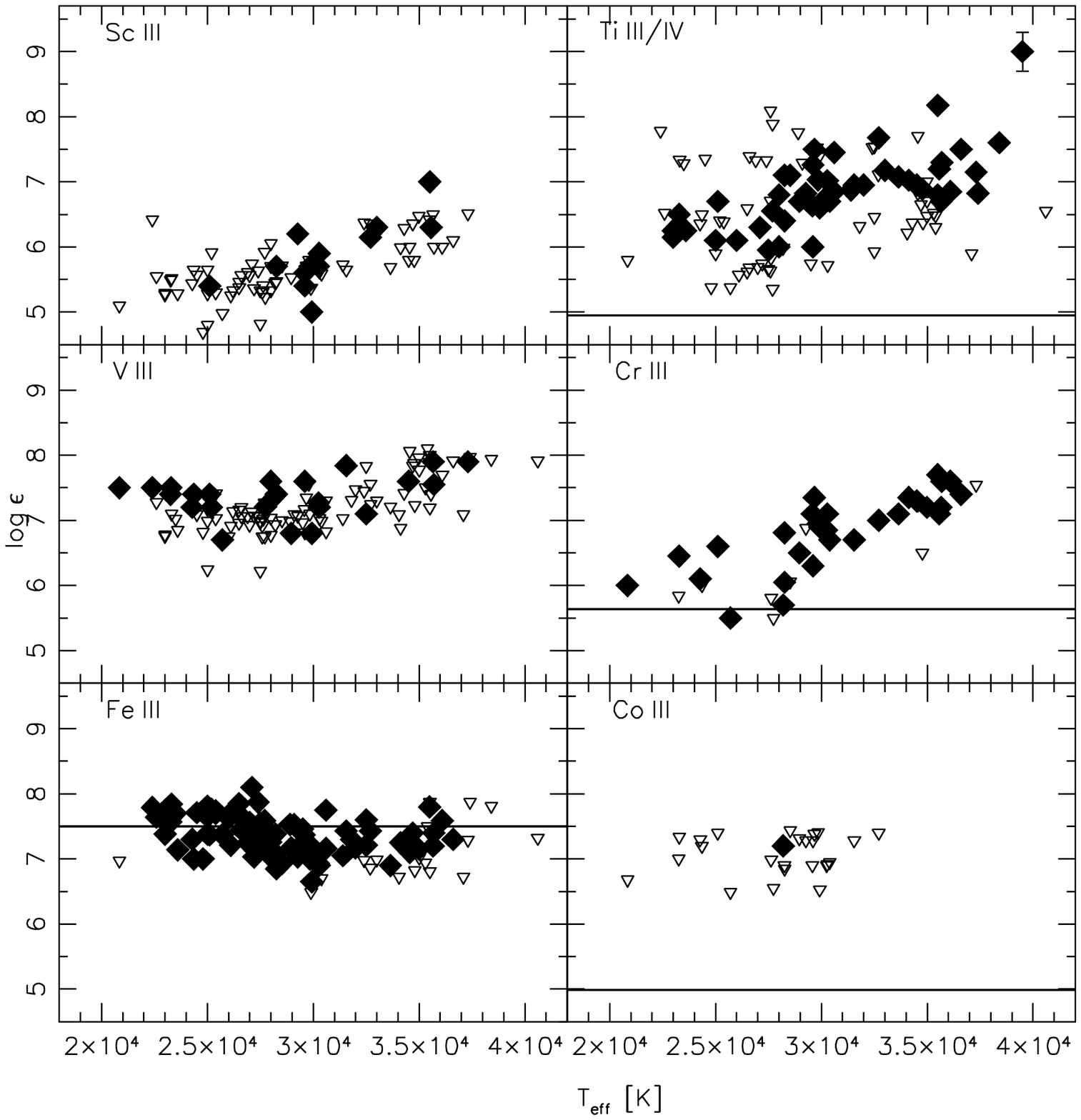}}
	\caption{Elemental abundances from scandium to cobalt plotted against effective temperature (see Fig.~\ref{abuns1}). In the case of scandium and vanadium the solar abundances are lower than $4.5\,{\rm dex}$ (see text)}
	\label{abuns3}
\end{center}
\end{figure*}

\section{Comparison with literature and systematic uncertainties \label{sec:lit}}

Important sources of systematic errors are discussed in Przybilla et al. (\cite{przybilla06}) and references therein. Enhanced metal abundances can cause significant line blanketing (e.g. O'Toole \& Heber \cite{otoole06}; Geier et al. \cite{geier07}), which can affect the temperature-pressure stratification of the atmosphere and therefore the atmospheric parameter determination. However, these parameters are used to construct the model spectra for measuring the metal abundances. Some lines are much more sensitive to a change in the atmospheric parameters than others. Especially for the heavier elements there is a severe lack of adequate atomic data, which may introduce systematic shifts of the derived abundances. NLTE (non-local thermodynamic equilibrium) effects which are neglected in this analysis become important especially at higher temperatures. But not all lines are equally affected by NLTE effects. In detailed analyses of main sequence B stars, some lines were found to behave well in LTE, while others can deviate in abundance by orders of magnitude when NLTE effects are taken into account (Nieva \& Przbilla \cite{nieva08}). Finally, microturbulence may lead to systematic trends in the abundances derived from single lines dependent on the line-strength. Due to the limited number of lines used in this study, it is not possible to measure this effect directly. However, more detailed analyses of a few stars in literature showed that in most cases microturbulence in sdB atmospheres is either negligible or small (e.g. Heber et al. \cite{heber00}; Edelmann \cite{edelmann03a}; Przybilla et al. \cite{przybilla06}).

In order to check whether the abundances determined with the pipeline approach are consistent with those derived by more detailed studies, we compared our results with independent determinations from literature. While we restrict ourselves to a limited set of pre-selected metal lines, other studies usually use all suitable lines in the spectra. The latter approach is more challenging and time consuming. 

Metal abundances of HD\,4539, PG\,1710$+$490, PG\,1627$+$027, PG\,1716$+$426, PHL\,457 and PG\,0101$+$039 have been determined by Fontaine et al. (\cite{fontaine06}) and Blanchette et al. (\cite{blanchette08}) using FUV-spectra obtained with the FUSE satellite. O'Toole \& Heber (\cite{otoole06}) performed a detail analysis of UV-spectra obtained with the HST/STIS spectrograph for Feige\,48, CD$-$24$^{\rm \circ}$731 and CPD$-$64$^{\rm \circ}$481. Optical high-resolution spectra were used to determine the abundances of PG\,1101$+$249 (Feige\,36, Edelmann et al. \cite{edelmann99}), HE\,1047$-$0436 (Napiwotzki et al. \cite{napiwotzki01}), TON\,S\,183 (Pereira \& Jeffery \cite{pereira08}), KPD\,2109$+$440, PG\,1219$+$534 (Heber et al. \cite{heber00}), HD\,205805 and Feige\,49 (Przybilla et al. \cite{przybilla06}). 

Fig.~\ref{abun_residuals} shows the differences of our abundances and the results from literature. Despite different wavelength ranges of the spectra, different atomic data and different methods to calculate the models (LTE or NLTE) the average scatter ($\pm0.2\,{\rm dex}$) is consistent with our estimate of the statistical uncertainties. In conclusion, the metal abundances determined using the pipeline method are in general consistent with the ones determined by more detailed approaches.

However, in a few cases the differences between our results and the results from literature are more significant ($>0.5\,{\rm dex}$). The same holds for the ionisation equilibria of some elements, which are very sensitive to the atmospheric parameters, especially the effective temperature. In most cases such mismatches can be  explained and a few of them are discussed now. 

The abundances of C~{\sc ii} and C~{\sc iii} differ by more than $0.6\,{\rm dex}$ in some stars (CPD$-$64$^{\rm \circ}$481, HE\,1047$-$0436) while the abundances of other elements in different ionisation stages are consistent within the error margins. Since the C~{\sc ii}/C~{\sc iii} abundances given in literature (Napiwotzki et al. \cite{napiwotzki01}; O'Toole \& Heber \cite{otoole06}) do not show such high deviations, we conclude that the limited number of lines used in our study leads to those systematic shifts. Especially the C~{\sc ii} lines at $4267\,{\rm \AA}$ are known to be very sensitive to NLTE effects (Nieva \& Przybilla \cite{nieva08}). In a pilot study, Przybilla et al. (\cite{przybilla06}) derived the carbon abundance from the C~{\sc ii} and C~{\sc iii} lines of HD\,205805 using LTE and NLTE models. The mismatch of $0.43\,{\rm dex}$ using the LTE approach consistent with our results could be significantly reduced to $0.17\,{\rm dex}$ using NLTE models.
However, the similar difference between the abundances derived from C~{\sc ii} and C~{\sc iii} lines in the hotter sdB Feige\,49, which is measured here as well, could not be reduced in this way. 

The most significant differences in the carbon ionisation equilibrium are seen in HE\,0101$-$2707 ($0.9\,{\rm dex}$) and HS\,2033$+$0821 ($1.2\,{\rm dex}$), where the abundances derived from the C~{\sc iii} lines are much higher than the ones derived from the C~{\sc ii} lines. The most likely explanation for these mismatches are pecularities in the line profiles discussed in Sect.~\ref{sec:strat}.

The abundances derived from N~{\sc ii}/N~{\sc iii} lines, Si~{\sc iii}/Si~{\sc iv} lines and S\,{\sc ii}/S\,{\sc iii} lines agree within the given error margins for most stars where both ionisation stages are present. Although the abundances of Ti~{\sc iii} and Ti~{\sc iv} agree within the given error margins in most stars, in some cases the difference can be as high as $1.4\,{\rm dex}$. NLTE effects or insufficient atomic data are the most likely reasons for these discrepancies.

\section{Metal abundances \label{sec:metal}}

Metal abundances of all stars are given in Table~A.1 and are plotted against effective temperature in Figs.~\ref{abuns1}, \ref{abuns2} and \ref{abuns3}. Atmospheric parameters and helium abundances are published in Lisker et al. (\cite{lisker05}), Geier et al. (\cite{geier12b}) and Paper I, the projected rotational velocities in Paper I and Paper II. 

\subsection{Carbon}

The observed C abundances derived from C~{\sc ii/iii} lines scatter from $-2.5\,{\rm dex}$ subsolar to solar (solar abundance $8.43\,{\rm dex}$). Three sdBs with $T_{\rm eff}>32\,000\,{\rm K}$ show supersolar abundances up to more than $+1.0\,{\rm dex}$. 

\subsection{Nitrogen and oxygen}

The N\,{\sc ii} abundances (solar abundance $7.83\,{\rm dex}$) range from $-1.0\,{\rm dex}$ to $+0.5\,{\rm dex}$ and do not show any trend with temperature. The O\,{\sc ii} abundances (solar abundance $8.69\,{\rm dex}$) range from $-2.0\,{\rm dex}$ to solar. At a temperature of $\simeq30\,000\,{\rm K}$ the average abundance is shifted by $-0.5$ with respect to the cooler stars in the sample. 

\subsection{Neon and magnesium}

The Ne\,{\sc ii} abundance scatters from $-1.5\,{\rm dex}$ to $+0.1\,{\rm dex}$ (solar abundance $7.93\,{\rm dex}$). A slight trend can be seen with temperature in the magnesium abundance (solar abundance $7.60\,{\rm dex}$), which ranges from $-1.5\,{\rm dex}$ to $-0.2\,{\rm dex}$.

\subsection{Aluminium and silicon}

A slight trend with temperature is present in the Al\,{\sc iii} abundance (solar abundance $6.45\,{\rm dex}$). Aluminium is enriched from $-1.5\,{\rm dex}$ to $0.0\,{\rm dex}$. The abundances from Si\,{\sc iii} as well as from Si\,{\sc iv} (solar abundance $7.51\,{\rm dex}$) show a large scatter between $-2.0\,{\rm dex}$ and $0.0\,{\rm dex}$ w. r. t. solar. Subdwarfs with strong silicon lines are present in the same temperature range as sdBs where only low upper limits can be given. At temperatures higher than $\simeq35\,000\,{\rm K}$ the mean silicon abundance drops by about $-1.0\,{\rm dex}$. 

\subsection{Phosphorus and sulfur}

A trend with temperature is present in the P\,{\sc iii} abundance (solar abundance $5.41\,{\rm dex}$). Phosphorus is enriched starting at a temperature of $T_{\rm eff}>28\,000\,{\rm K}$ from $-0.5\,{\rm dex}$ to $+1.0\,{\rm dex}$. The two stars HE\,2307$-$0340 and HE\,0539$-$4246 with temperatures of $\simeq23\,000\,{\rm K}$ show lines at a wavelength of P\,{\sc iii} $4080.089\,{\rm \AA}$. Although no possible blends were found in line lists, these lines may be misidentifications since the derived abundances ($5.90\,{\rm dex}$) seem to be too high to fit in the overall trend. The S\,{\sc ii} and especially S\,{\sc iii} abundances (solar abundance $7.12\,{\rm dex}$) scatter between $-1.5\,{\rm dex}$ and $+1.0\,{\rm dex}$. 

\subsection{Argon, potassium and calcium}

The Ar abundance increases with temperature from solar to $+1.8\,{\rm dex}$ (solar abundance $6.40\,{\rm dex}$). This trend has not been reported in prior analyses. Potassium has not been discovered in sdB atmospheres so far (solar abundance $5.03\,{\rm dex}$). Similar to argon, the K abundance increases with temperature from $+0.7\,{\rm dex}$ to $+3.0\,{\rm dex}$. Ca\,{\sc iii} (solar abundance $6.34\,{\rm dex}$) is present at temperatures higher than $T_{\rm eff}>29\,000\,{\rm K}$. The abundances scatter from $+1.0\,{\rm dex}$ to $+2.5\,{\rm dex}$. Ca\,{\sc ii} was not included in our analysis because the most prominent lines are usually blended with interstellar lines. 

\subsection{Scandium, titanium, vanadium and chromium}

Sc\,{\sc iii} (solar abundance $3.15\,{\rm dex}$) is strongly enriched and its abundance increases with temperature from $+2.0\,{\rm dex}$ to $+4.0\,{\rm dex}$. Ti\,{\sc iii} (solar abundance $4.95\,{\rm dex}$) is enriched and scatters from $+1.0\,{\rm dex}$ to $+3.0\,{\rm dex}$.  V\,{\sc iii} (solar abundance $3.93\,{\rm dex}$) is highly enriched independent of the temperature ranging from $+2.0\,{\rm dex}$ to almost $+4.0\,{\rm dex}$. The Cr\,{\sc iii} abundance (solar abundance $5.64\,{\rm dex}$) increases with temperature from $+0.0\,{\rm dex}$ to $+2.0\,{\rm dex}$.

\subsection{Iron, cobalt and zinc}

The Fe\,{\sc iii} abundance (solar abundance $7.50\,{\rm dex}$) is constant ranging from $-0.7\,{\rm dex}$ to $+0.5\,{\rm dex}$. For cobalt (solar abundance $4.99\,{\rm dex}$) and zinc (solar abundance $4.56\,{\rm dex}$) only upper limits could be given (zinc is not shown in Fig.~\ref{abuns3} because upper limits could be derived from a single line only). These limits allow high enrichments of these elements up to $+2.0\,{\rm dex}$, which is consistent with the results from UV-spectroscopy (O'Toole \& Heber \cite{otoole06}; Chayer et al. \cite{chayer06}; Blanchette et al. \cite{blanchette08}).

\begin{figure*}[t!]
\begin{center}
	\resizebox{9cm}{!}{\includegraphics{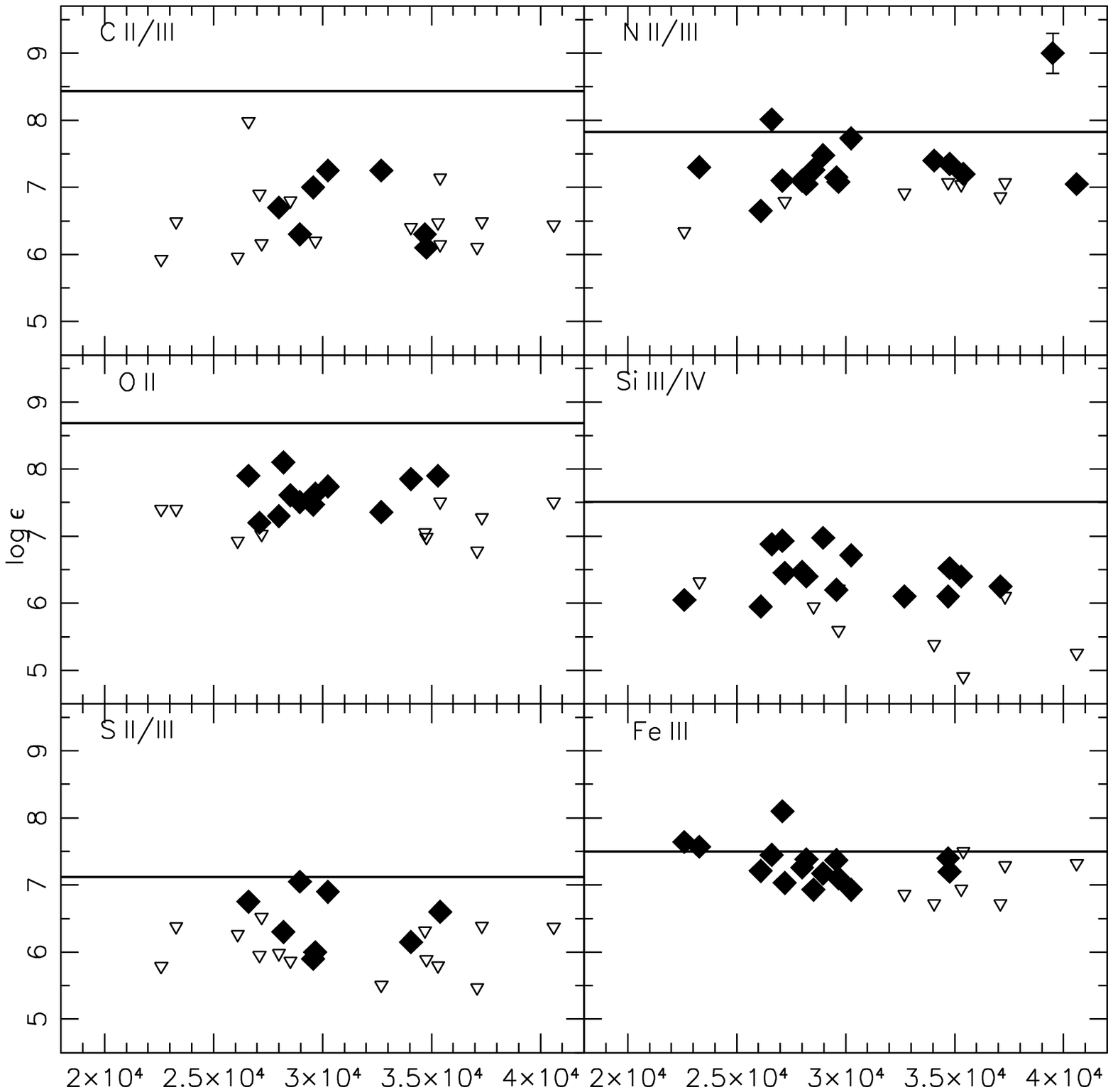}}
	\resizebox{9cm}{!}{\includegraphics{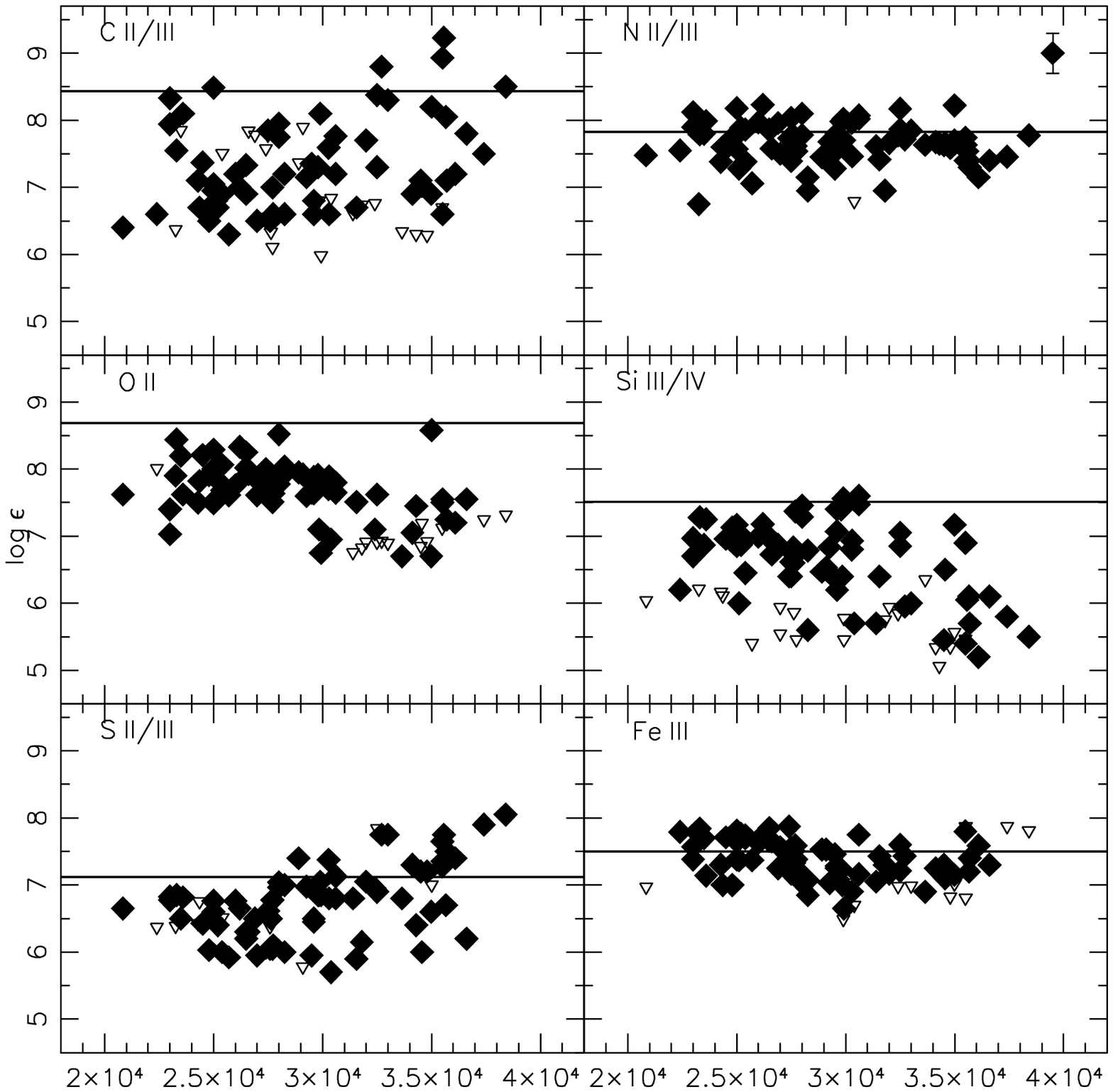}}
	\caption{Abundances of selected elements plotted against effective temperature (see Fig.~\ref{abuns1}). {\it Left panel} Only sdBs belonging to the lower He-sequence are plotted. {\it Right panel} Only sdBs belonging to the upper He-sequence are plotted.}
	\label{abuns_helow}
\end{center}
\end{figure*}

\section{Abundance patterns of sdB sub-populations \label{sec:sub}}

Besides apparently single sdB stars our sample contains several pulsating sdBs and RV-variable systems in close binaries. These stars may have formed differently or - in the case of the pulsating sdBs - may be in a distinct evolutionary phase. It is instructive to compare the metal abundance patterns of these sub-populations and search for differences, which may help to clarify these issues.

Our sample contains 38 RV-variable sdBs in single-lined, close spectroscopic binary systems. Aznar Cuadrado \& Jeffery (\cite{aznar02}) argued that the helium content in such close binary sdBs may be higher than in single stars, because tidal effects might lead to mixing in the stellar atmosphere. According to this scenario, the metal abundances should be affected as well. However, we showed that the helium content in sdB atmospheres is not affected in this way (Geier et al. \cite{geier12b}). 

There is also no significant difference between the metal abundance patterns of sdBs in close binaries and apparently single stars. In conclusion, moderate tidal influence of close companions does not change the abundances in sdB atmospheres. In the most extreme cases, where the sdB has been spun up to very high rotational velocities of the order of $100\,{\rm km\,s^{-1}}$ (e.g. KPD\,1930$+$2752, Geier et al. \cite{geier07}; EC\,22081$-$1916, Geier et al. \cite{geier11a}), it was not yet possible to determine the metal abundances, because the broadening of the spectral lines is too strong. Such objects may still show peculiar abundance patterns. 

Since the driving mechanism of pulsations in sdB stars is thought to be an enrichment of iron in the outer layers of the star (Charpinet et al. \cite{charpinet97}), the metal abundances of pulsating sdBs have been determined to search for peculiarities. However, the abundance patterns of pulsating sdBs turned out to be the same as the ones of non-pulsating comparison stars with similar atmospheric parameters (Heber et al. \cite{heber00}; O'Toole \& Heber \cite{otoole06}; Blanchette et al. \cite{blanchette08}). 

Our sample includes the short-period pulsators Feige\,48 (Koen et al. \cite{koen98}), KPD\,2109$+$440 (Bill\`{e}res et al. \cite{billeres98}), PG\,1219$+$534 (O'Donoghue et al. \cite{odonoghue99}) and HE\,0230$-$4323 (Kilkenny et al. \cite{kilkenny10}) as well as the long-period pulsators PG\,1627$+$017 (Green et al. \cite{green03}), LB\,1516 (Koen et al. \cite{koen10}), JL\,82 (Koen \cite{koen09}), PHL\,44 (Kilkenny et al. \cite{kilkenny07}), PHL\,457 (Blanchette et al. \cite{blanchette08}), PG\,1716$+$426 (Green et al. \cite{green03}) and PG\,0101$+$039 (Randall et al. \cite{randall05}). Consistent with the results of Heber et al. (\cite{heber00}), O'Toole \& Heber (\cite{otoole06}) and Blanchette et al. (\cite{blanchette08}) no differences have been found in the metal abundance patterns of these stars with respect to the rest of the sample.

\section{Helium and metal abundances \label{sec:helium}} 

Edelmann et al. (\cite{edelmann03b}) discovered a correlation between the effective temperatures of sdB stars and their helium abundances. Hotter sdB stars tend to show higher enrichments of helium in some cases reaching or even exceeding the solar abundance. However, the scatter in helium abundance is high and two distinct sequences appear in the $T_{\rm eff}-\log{y}$-diagram. These results have been confirmed by Geier et al. (\cite{geier12b}). About $75\%$ of the field sdB population belongs to the upper sequence, while $\simeq25\%$ form the lower sequence, which is offset by about $2\,{\rm dex}$. 

It is instructive to compare the metal abundances of sdBs belonging to the two helium sequences (see Fig.~\ref{abuns_helow}). Most species are not significantly affected by the difference in helium abundance. For nitrogen, oxygen, iron and all other metals not shown in Fig.~\ref{abuns_helow} there are no obvious differences between the two sub-samples. The silicon abundances of the helium-poor stars seem to be somewhat lower than for the helium-rich ones. However, this impression is most likely created by the different sizes of the samples. 

In contrast, significant differences are visible for the carbon and sulfur abundances. All sdBs on the lower helium sequence have carbon abundances of $-1.0$ or less with respect to the solar value, while the stars with more helium show a large scatter in carbon abundance up to supersolar values. The sulfur abundances behave in a similar way. While the helium-poor sample shows subsolar abundances, sulfur can be enriched to supersolar values in the helium-rich sample. 

\section{Other trends \label{sec:other}}

It has been noted by several authors (e.g. Lamontagne et al. \cite{lamontagne85}, \cite{lamontagne87}; O'Toole \& Heber \cite{otoole06}) that the silicon abundance in sdB stars appears to drop sharply at $T_{\rm eff}>32\,000\,{\rm K}$. Fractionated winds have been invoked to explain this strange observation (Unglaub \cite{unglaub08}). The aluminium abundance were proposed to behave in a similar way. As can be clearly seen in Figs. \ref{abuns1} and \ref{abuns2}, this assumption has to be dropped. Silicon as well as aluminium are present all over the temperature range of sdBs. 

O'Toole \& Heber (\cite{otoole06}) also reported a possible anti-correlation between iron and the other heavy elements. In our sample no such trend is visible for the Ti and Ca abundances. 

Looking at high resolution spectra of a sample of sdBs it is striking that most of them show a lot of metal lines, while some don't show any metal lines at all. At first instance one could argue, that there exist two different populations of sdBs with different metallicities. Heber \& Edelmann (\cite{heber04}) discovered three hot sdBs showing lots of metal lines (e.g. PG\,0909$+$276). They were subsequently named "super-metal-rich" sdBs because their metal abundances seemed to be exceptionally high. 

Our results provide a natural explanation for this effect, since such an enrichment is found to be quite normal for sdBs in the high temperature range. All stars with no metal lines are lying at the hot end of the EHB with temperatures higher than $T_{\rm eff}>33\,000\,{\rm K}$. Although they show no lines, the derived upper limits are consistent with the adundance measurements of hot sdBs with metal lines. At these temperatures the optical lines are becoming so weak that they are only observable in high S/N spectra. The effect can be seen in Fig. \ref{abuns1}, where the oxygen and magnesium abundance measurements partly turn into upper limit estimates at high temperatures. Spectra of slightly poorer quality appear to be metal free, but they are not. This selection effect illustrates the limitation of optical spectroscopy for metal abundance analyses in hot stars. Therefore UV spectra are needed to study the strong metal lines of higher ionisation stages in hotter sdO/B or sdO stars.

\section{Extreme horizontal branch versus blue horizontal branch}\label{sec:bhb}

\begin{figure}[h!]
\begin{center}
	\resizebox{7.5cm}{!}{\includegraphics{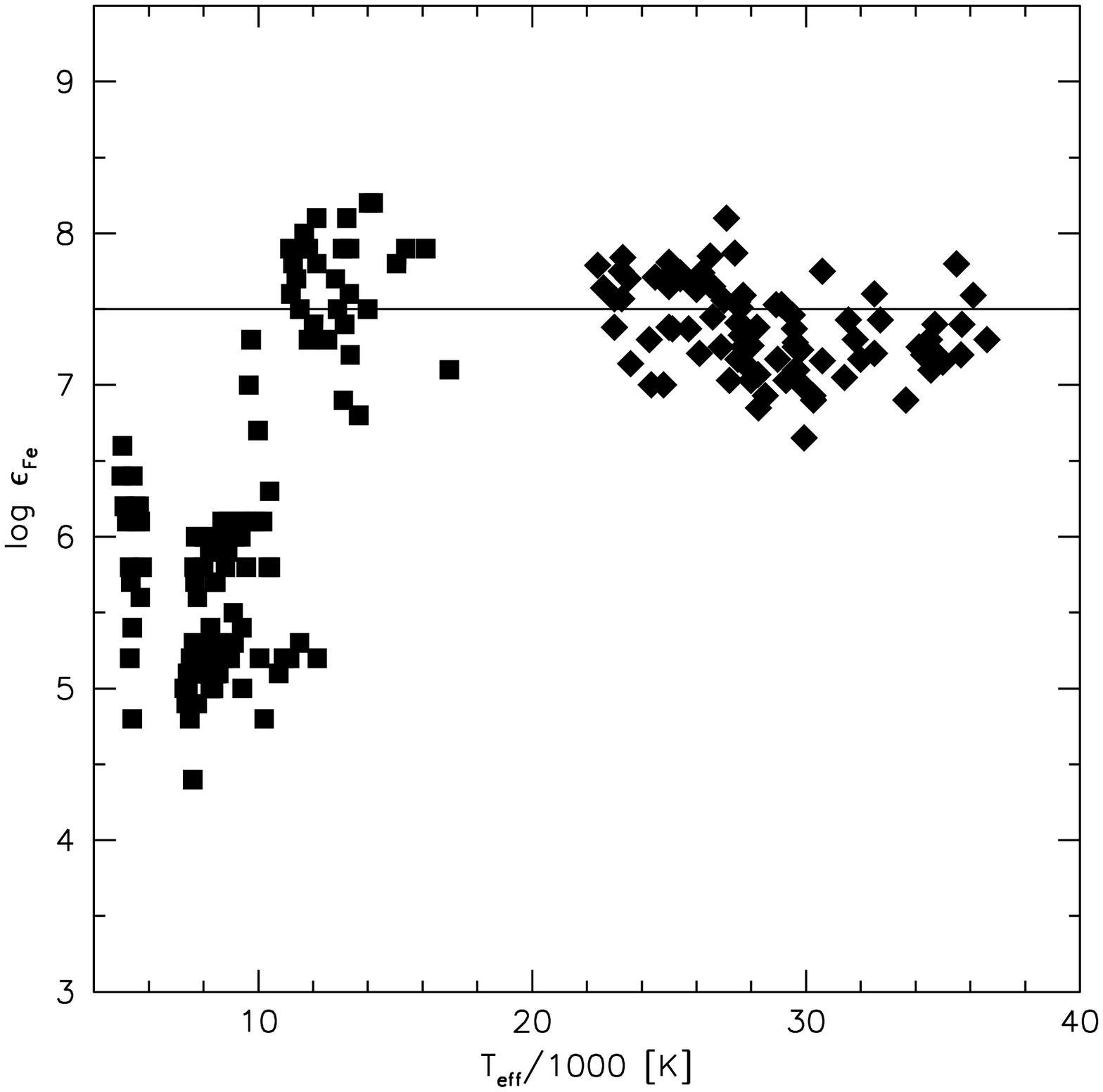}}
	\resizebox{7.5cm}{!}{\includegraphics{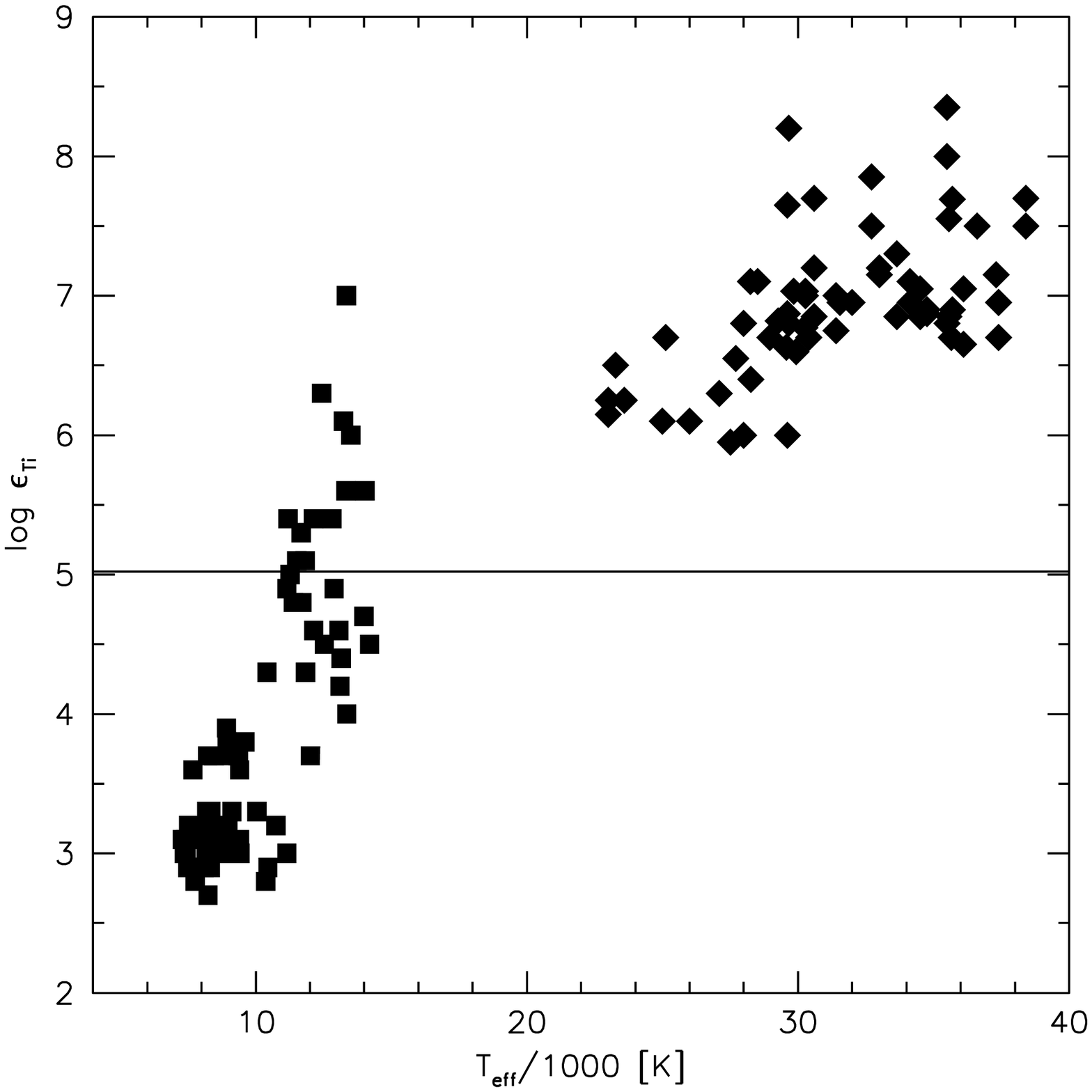}}
\end{center}
\caption[Iron and titanium abundances of sdBs and BHBs plotted against effective temperature.]{{\it Left panel} Iron abundance is plotted against effective temperature. The filled diamonds mark the results from the sdBs. The filled rectangles mark the combined results for BHB stars from seven GCs and the field (Behr et al. \cite{behr03a,behr03b}; Fabbian et al. \cite{fabbian05}). {\it Right panel} Titanium abundance is plotted against effective temperature. The filled diamonds mark the results from the sdBs. The filled rectangles mark the combined results for BHB stars from eight GCs and the field (Behr et al. \cite{behr03a,behr03b}; Fabbian et al. \cite{fabbian05}; Pace et al. \cite{pace06}).}
\label{bhb_diff}
\end{figure}

Quantitative spectral analyses of BHB stars have been published by Behr et al. (\cite{behr03a, behr03b}), Fabbian et al. (\cite{fabbian05}) and Pace et al. (\cite{pace06}) and it is worthwhile to compare them to our results for EHB stars. In Fig.~\ref{bhb_diff} the abundances of iron and titanium are plotted against effective temperature. Abundances of all BHB stars are plotted together, although they are derived from stars coming from very different populations. Eight different GCs as well as field stars are put together here.\footnote{The gap in temperature between the BHB and EHB stars is a known, but yet unexplained feature, which is also observed in two-colour diagrams of field blue halo stars and GCs (Newell \cite{newell73}; Geier at al. \cite{geier11b} and references therein).} 
At first glance this makes absolutely no sense, because stars from different chemical environments should have different abundances. This would result in a scattered plot from which no relevant information could be derived.

However, Fig.~\ref{bhb_diff} proves that this is not the case. Although there is a high scatter in the iron abundance at low temperatures, all stars hotter than about $11\,500\,{\rm K}$ end up at almost solar abundance with a significantly lower scatter (Fig.~\ref{bhb_diff}, upper panel). As soon as diffusion sets in, the atmospheres of stars from different populations become similar regardless of the primordial abundances. The distribution of sdBs now shows that the iron abundance remains indeed saturated at this value up to temperatures of $40\,000\,{\rm K}$. This plot clearly illustrates that the abundance of iron in EHB and the hottest BHB stars is not "solar" for reasons of star formation and stellar evolution. This abundance reflects the surface concentration of iron caused by an interplay of gravitational settling and radiative levitation, which becomes saturated in stars hotter than $11\,500\,{\rm K}$, and is just by chance "solar". This result is in perfect agreement with diffusion models (Michaud et al. \cite{michaud08}, \cite{michaud11}).
 
A similar behaviour is predicted by Michaud et al. (\cite{michaud08}) for the titanium abundance. In BHB stars only a rise of the titanium abundance can be observed, which is more continuous than in the case of iron. Adding the sdBs one can see that the abundance becomes saturated at an effective temperature of about $30\,000\,{\rm K}$ and an abundance of roughly 100 times solar (Fig.~\ref{bhb_diff}, lower panel). This behaviour proves in a most convincing way, that heavy elements in EHB and hot BHB stars are enriched by radiative levitation. 

\section{Diffusion at work along the horizontal branch \label{sec:lev}}

The metal abundances of the sdBs in our sample show a pattern, which could not been seen that clearly before. While the light elements carbon, nitrogen, oxygen and neon are not affected by higher temperatures, most heavier elements from aluminium to chromium become enriched in hotter atmospheres, the high scatters in the silicon and sulfur abundance being interesting exceptions. Iron, on the other hand, becomes saturated at solar abundance in HB star atmospheres as soon as the effective temperature exceeds $11\,500\,{\rm K}$. Consistent with that the hotter sdB stars all have "solar" iron abundances.

These results are in reasonable agreement with theoretical calculations performed by Michaud et al. (\cite{michaud08}, \cite{michaud11}). The authors calculated full evolutionary models, including the effects of diffusion and radiative acceleration for different primordial metallicities. The evolution on the HB was followed for the first $32\,{\rm Myr}$, which is about one third of the typical lifetime of core helium-burning stars. Their calculations of the abundance anomalies only depend on the mass mixed by turbulence at the surface. Other effects like mass-loss via stellar winds have not been taken into account. Michaud et al. (\cite{michaud11}) used the observed iron abundances in sdB stars mostly taken from a preliminary version of the dataset presented here to fix the mixed surface mass to $\simeq10^{-7}\,M_{\rm \odot}$. 

The time evolution of the surface abundances was then calculated for the elements C, N, O, Ne, Mg, Al, Si, P, S, Ar, K, Ca, Ti, Cr and Fe. In order to compare our measurements with these curves we can restrict ourselves to models with solar metallicity on the main sequence (Michaud et al. \cite{michaud11}). Since the sdBs in our sample are rather bright field stars, we can assume that most of them originate from the Galactic thin disk with only minor contributions from the thick disk or the halo (e.g. Altmann \& de Boer \cite{altmann00}). 

In general, the models of Michaud et al. (\cite{michaud11}) match the observations quite well. The predicted enrichments or depletions are mostly of the same order as the observations. This is the first time that the complete abundance patterns of sdB stars could be modelled in a quantitative way. However, the models do neither predict the highly supersolar enrichments of C, S, P, K and Ar with temperature nor the large scatter in the C, Si and S abundances. 

Both the high enrichments and the large scatter of certain abundances may partly be caused by age effects. Michaud et al. (\cite{michaud11}) pointed out that their evolutionary calculations only cover a part of the lifetime on the EHB. There are no clear correlations between the metal abundances and the ages of the stars derived from their position in the $T_{\rm eff}-\log{g}$-diagram. However, such a trend may be washed out by the uncertainties in the atmospheric parameters and the abundances. Furthermore, mass loss via stellar winds, which was not accounted for, might play a role as well (e.g. Unglaub \cite{unglaub08} and references therein). However, a closer inspection of the data makes an alternative explanation more likely.

\begin{figure}[t!]
	\resizebox{\hsize}{!}{\includegraphics{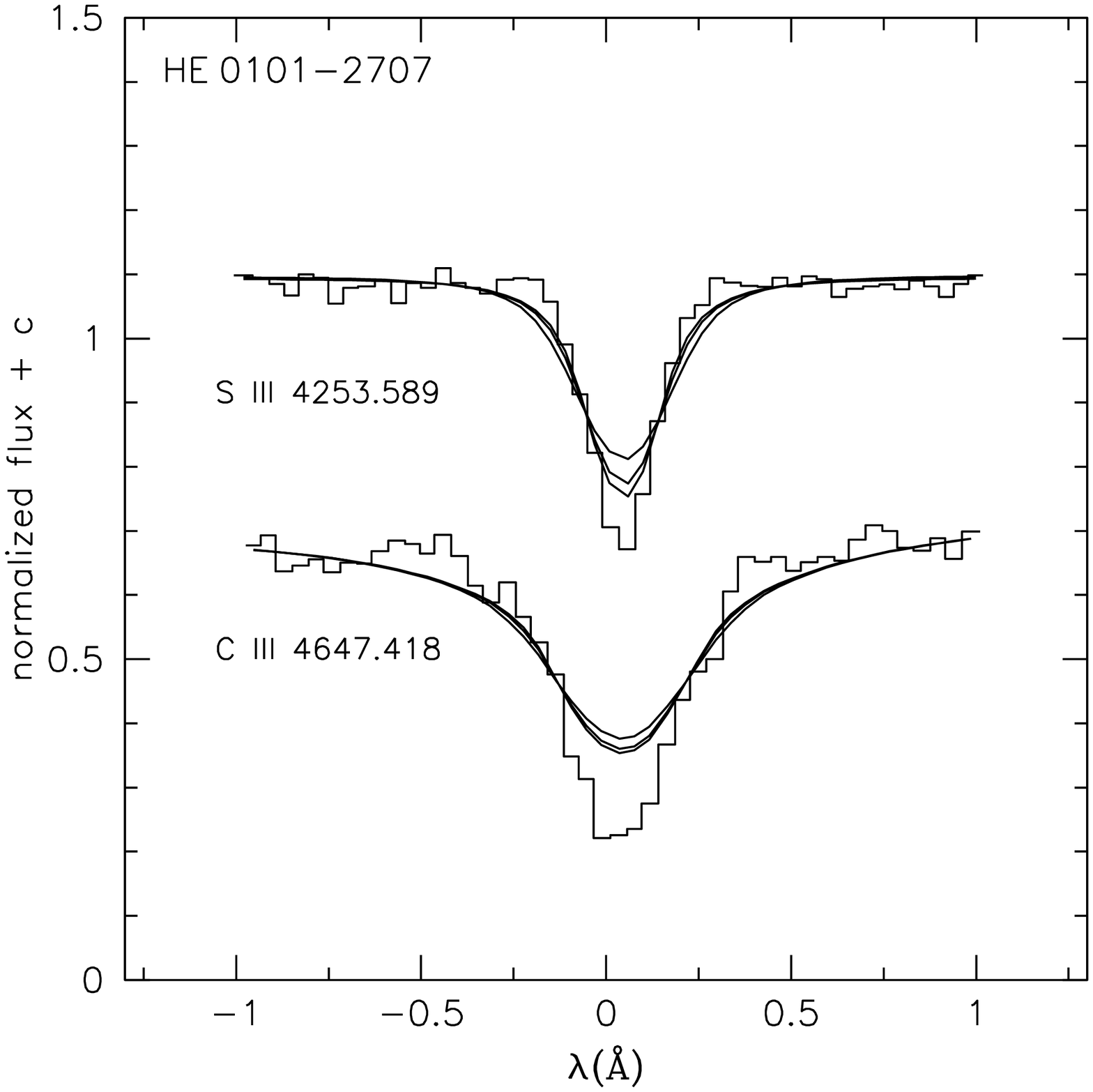}}
	\resizebox{\hsize}{!}{\includegraphics{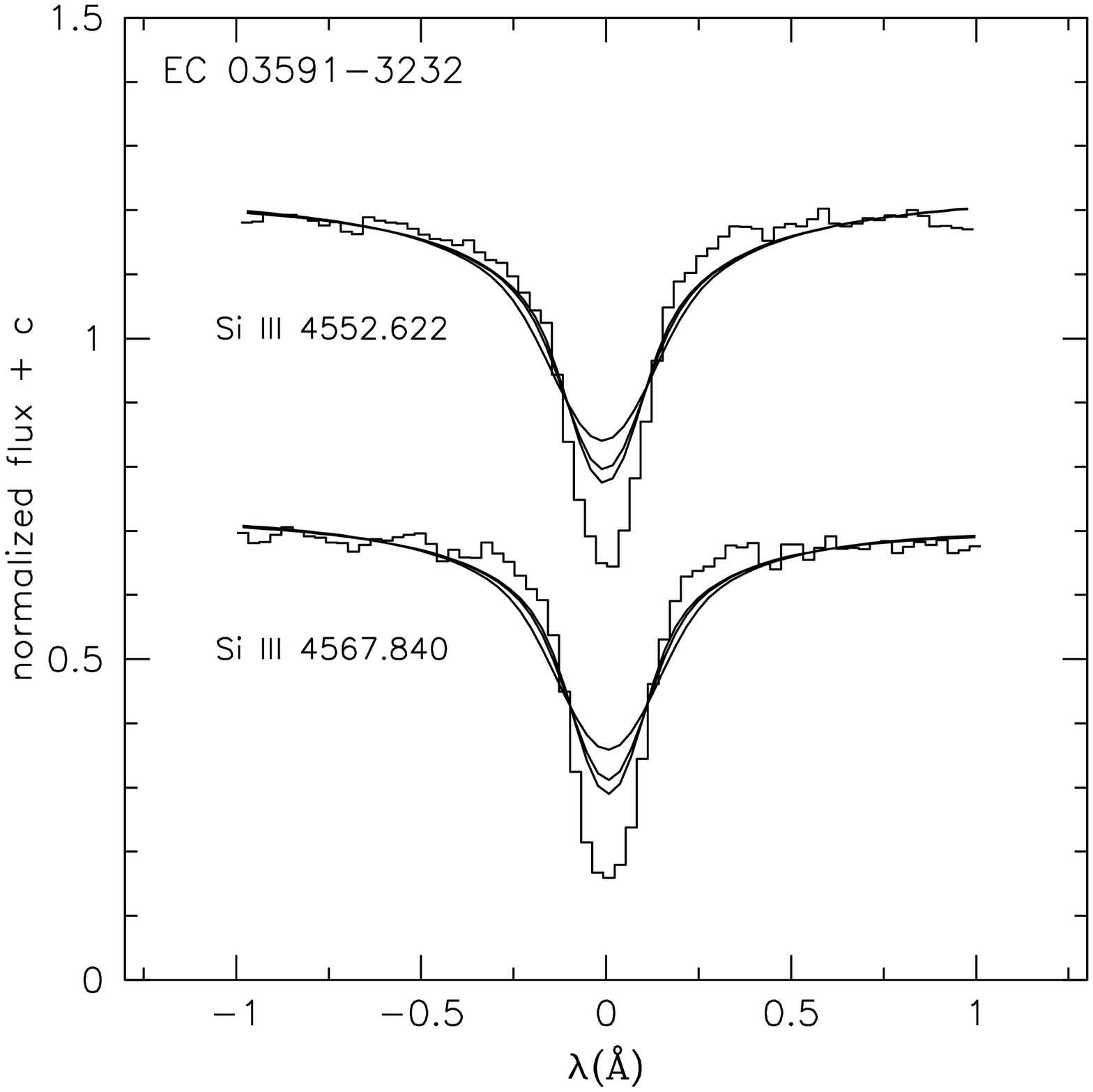}}
	\caption{{\it Upper panel} Fits of models with different rotational broadening (solid lines, $0,5,10\,{\rm km\,s^{-1}}$ respectively) to an S\,{\sc iii} and a C\,{\sc iii} line (solid histograms) of HE\,0101$-$2707. {\it Lower panel} Similar fits of models to Si\,{\sc iii}-lines of EC\,03591$-$3232. It can be clearly seen that the synthetic line profiles do not match the observed ones at all. Examples for normal fits to metal lines are given in Fig.~\ref{quality}.}
	\label{strati}
\end{figure}

\section{Peculiar line profiles and stratification \label{sec:strat}}

While the enrichments of P, K, Ar (and maybe also S at the hot end) follow a general trend with temperature, which seems to be an indicator of radiative levitation, the supersolar enrichment of some stars in C is hard to explain in this framework, because the other light elements N, O, Ne and Al are depleted with respect to their solar abundances. 

Since carbon enrichment is also common in the helium-rich sdBs (Ahmad \& Jeffery \cite{ahmad03}; Naslim et al. \cite{naslim10}) and sdOs (Str\"oer et al. \cite{stroer07}; Hirsch \& Heber \cite{hirsch09}) and the helium abundances of hot sdBs with carbon enrichment are among the highest in our sample, a connection seems to be likely. Alternative formation scenarios like the late hot flasher (Sweigart \cite{sweigart97b}; Lanz et al. \cite{lanz04}) do indeed predict an enrichment of helium and carbon. 

In the case of helium enrichment prior to the subdwarf stage Groth et al. (\cite{groth85}) predict a broad subsurface convection zone, which should be able to mix nuclear processed material in the atmosphere and therefore counteract gravitational settling. Of course, this mechanism would have to ensure the enrichment of He and C without destroying the general abundance pattern caused by diffusion, which might require some finetuning. Diffusion might also lead to a slow transformation of He-rich subdwarfs into He-deficient ones (Miller Bertolami et al. \cite{miller08}). The discovery of sdBs with intermediate He abundances right in between the normal sdBs and the He-rich ones may be consistent with such a scenario (Ahmad et al. \cite{ahmad07}; Naslim et al. \cite{naslim11}; Naslim et al. \cite{naslim12}).

However, we found observational evidence that yet another unaccounted effect may be responsible for the most extreme chemical peculiarities in sdB atmospheres. Looking at the spectra of our sample in detail, the shape of the strongest metal lines in some stars (CPD$-$64$^{\rm \circ}$481, EC\,03591$-$3232, EC\,05479$-$5818, Feige\,38, HD\,205805, HE\,0101$-$2707, PG\,0133$+$144, PG\,1303$+$097, PG\,1710$+$490, PG\,1743$+$477) was found to be peculiar (see Fig. \ref{strati}). The lines are too sharp to be fitted with synthetic models even if the rotational broadening is set to zero. Especially the line wings are much too weak with respect to the depths of the cores. NLTE effects cannot cause such line shapes and the weaker metal lines don't show similar profiles. 

Using five spectra of Feige\,38 taken within a timespan of $1.8\,{\rm yr}$ we checked whether the shape of the peculiar lines changes. The lines do not show any significant differences in depth and shape. Time dependent effects like star spots can therefore be excluded.

The most significant pecularities are seen in the strongest Si\,{\sc iii} lines of EC\,03591$-$3232 (see Fig. \ref{strati}, lower panel), PG\,0133$+$114, Feige\,38 and PG\,1710$+$490. Most remarkably all of these stars belong to the rare class of sdBs that also show an enrichment in $^{3}$He (Geier et al. \cite{geier12b}).  

Michaud et al. (\cite{michaud11}) pointed out that additional element separation may occur in the stellar atmosphere between the bottom of the mixed mass ($\simeq10^{-7}\,M_{\rm \odot}$) and the surface. Such effects are not included in their models. The $^{3}$He anomaly observed in sdB stars with temperatures around $30\,000\,{\rm K}$ is not predicted by the model of Michaud et al. (\cite{michaud11}) either and should therefore not be present in the mixed zone. Nevertheless, it is clear that the settling of almost all $^{4}$He must be caused by diffusion, which can then only happen in the outermost atmosphere. 

Diffusion in the photosphere should lead to vertical stratification of the metal abundances. Observational evidence for such a stratification of Fe in the atmospheres of BHB stars has been found (Khalack et al. \cite{khalack07, khalack10}; LeBlanc et al. \cite{leblanc09, leblanc10}). How is the shape of the spectral lines affected by vertical stratification? If radiative levitation leads to an abundance gradient in the atmosphere, the wings of a spectral line should become weaker, because of less absorption at the bottom of the atmosphere. This not only explains the peculiar shape of the lines, but also the fact that the weak lines are not affected. Those lines originate from deeper layers, where the gradient is less pronounced. Together with the $^{3}$He anomaly the peculiar line shapes provide evidence for diffusion processes above the mixed zone in the atmospheres of sdBs. 

Such effects may also provide a reasonable explanation for the high scatter of the C, Si and S abundances as well as mismatches in the ionisation equilibria (see Sect.~\ref{sec:lit}), since peculiar line shapes have also been found for strong C\,{\sc iii} and S\,{\sc iii} lines in the spectra of hot sdBs (e.g. HE\,0101$-$2707, see Fig.~\ref{strati}). Given that these peculiarities are only detectable in the most extreme cases or in data of high quality, stratification may be rather common in sdB atmospheres. It is therefore possible that the highest enrichments not predicted by the models of Michaud et al. (\cite{michaud11}) are caused by radiative levitation in the stellar atmosphere rather than the mixed zone beneath. Of course, this hypothesis needs to be tested further.

\section{Conclusion \label{sec:con}}

A combination of diffusion both in a mixed zone close to the surface and in the stellar atmosphere on top of it seem to be sufficient to explain most of the results presented here. However, different formation scenarios, which lead to different amounts of helium in the atmosphere may still leave some imprint on the final abundance patterns. The connection between sdBs with hydrogen-rich atmospheres, He-sdBs and intermediate objects remains as unclear as the reason for the differences in the abundance patterns of the two distinct helium populations (see Sect.~\ref{sec:helium}). Our lack of understanding the details of diffusion in sdBs is very well illustrated by the extremely peculiar intermediate He-sdB LS\,IV$-$14$^{\rm \circ}$116, whose atmosphere is highly enriched in strontium, yttrium and zirconium (Naslim et al. \cite{naslim11}). The reason for this is unclear and the very well detected spectral lines of these elements are not present in any of the sdBs of our sample. 

Furthermore, as pointed out by Hu et al. (\cite{hu11}), the physical mechanism necessary to mix the surface in the way needed to reproduce the observed abundances is still unknown. Mass loss through stellar winds may play a role, but Hu et al. (\cite{hu11}) also showed that the mass loss rates must be lower than $\simeq10^{-15}\,M_{\rm \odot}\,{\rm yr^{-1}}$ to allow sdBs to pulsate. Since the abundance patterns of sdB pulsators do not differ from the ones of stable sdBs, one would expect similar mass loss rates. But even very weak and fractionated winds might affect the atmospheric abundances (Unglaub \cite{unglaub08}).

Although in most cases unstratified photospheres reproduce the observed line profiles well, the discovery of spectral lines with peculiar shapes possibly caused by vertical stratification  might pose a new challenge for atmospheric models of sdB stars. Besides NLTE effects (Przybilla et al. \cite{przybilla06}) and enhanced line blanketing by metals enriched due to diffusion (O'Toole \& Heber \cite{otoole06}), next generation models might have to take vertical stratification into account. Especially the strong resonance lines in the UV should be affected and due to crowding and blending in this region such effects might not be obvious.

\vspace{1cm}
\begin{acknowledgements}
Based on observations at the Paranal Observatory of the European 
Southern Observatory for programmes number 165.H-0588(A), 167.D-0407(A), 071.D-0380(A) and 072.D-0487(A).
Based on observations at the La Silla Observatory of the 
European Southern Observatory for programmes number 073.D-0495(A), 074.B-0455(A), 076.D-0355(A), 077.D-0515(A) and 078.D-0098(A).
Based on observations collected at the Centro Astron\'omico Hispano Alem\'an (CAHA) at Calar Alto, operated jointly by the Max-Planck Institut f\"ur Astronomie and the Instituto de Astrof\'isica de Andaluc\'ia (CSIC). Some of the data used in this work were 
obtained at the Hobby-Eberly Telescope (HET), which is a joint project of the
 University of Texas at Austin, the Pennsylvania State University, Stanford 
 University, Ludwig-Maximilians-Universit\"at M\"unchen, and 
 Georg-August-Universit\"at G\"ottingen, for programmes number UT07-2-004 and UT07-3-005.
Some of the data presented here were obtained at the W.M. Keck Observatory, which is operated as a scientific 
 partnership among the California Institute of Technology, the University of California, and the National Aeronautics and Space Administration. The 
   Observatory was made possible by the generous financial support of the W.M. Keck Foundation.
S.~G. was supported by the Deutsche Forschungsgemeinschaft under grant 
He~1356/49-1. I would like to thank U. Heber, H. Edelmann, N. Przybilla, M. F. Nieva, K. Unglaub and I. Bues for their comments and fruitful discussions. 
\end{acknowledgements}

\begin{appendix}
\section{Metal abundances}
\onecolumn
\begin{landscape}
\begin{longtable}{lllllllllll}
\hline
\noalign{\smallskip}
Object              & $T_{\rm eff}$\,[K]      & He\,{\sc i/ii} & C\,{\sc ii}   & C\,{\sc iii}  & N\,{\sc ii}   & N\,{\sc iii} & O\,{\sc ii}    & Ne\,{\sc ii}  & Mg\,{\sc ii}  \\
\noalign{\smallskip}
\hline
\noalign{\smallskip}
HE\,0151$-$3919          & $20841$            & $9.93$         & $6.40$        &  $-$          & $7.48\pm0.13$ & $-$          & $7.62\pm0.26$  & $<8.15$       & $6.10$         \\
EC\,21494$-$7018         & $22400$            & $8.34$         & $6.60$        &  $-$          & $7.55\pm0.39$ & $-$          & $<8.01$        & $-$           & $<5.74$        \\
EC\,15103$-$1557         & $22600$            & $8.01$         & $<5.93$       &  $-$          & $<6.43$       & $-$          & $<7.40$        & $-$           & $6.00$         \\
EC\,11349$-$2753         & $23000$            & $9.87$         & $8.16\pm0.30$ & $8.50\pm0.14$ & $8.12\pm0.19$ & $-$          & $7.40\pm0.20$  & $-$           & $6.30$         \\
HD\,4539                 & $23000$            & $9.70$         & $8.08\pm0.23$ & $7.80$        & $7.90\pm0.16$ & $<8.34$      & $7.03\pm0.31$  & $<8.18$       & $6.60$         \\
HE\,2307$-$0340          & $23260$            & $8.35$         & $<6.37$       & $-$           & $6.75\pm0.07$ & $-$          & $7.90$         & $8.10\pm0.28$ & $6.30$         \\
HE\,0539$-$4246          & $23279$            & $8.09$         & $<6.49$       & $-$           & $7.30\pm0.22$ & $-$          & $<7.40$        & $<8.05$       & $6.00$         \\
EC\,14345$-$1729         & $23300$            & $9.41$         & $7.55\pm0.35$ & $<8.93$       & $7.79\pm0.15$ & $-$          & $8.44\pm0.27$  & $-$           & $6.50$         \\
PG\,1627$+$017$^{\rm r,l}$ & $23500$            & $9.20$         & $<7.85$       & $-$           & $7.80\pm0.12$ & $-$          & $8.20\pm0.08$  & $-$           & $6.60$         \\
PG\,1432$+$004           & $23600$            & $9.67$         & $8.10\pm0.25$ & $<8.51$       & $7.99\pm0.23$ & $-$          & $7.62\pm0.33$  & $-$           & $6.70$         \\
HE\,2208$+$0126$^{\rm r}$ & $24277$            & $9.02$         & $7.10$        & $-$           & $7.38\pm0.26$ & $<7.08$      & $7.50\pm0.26$  & $<8.10$       & $5.90$         \\
GD\,687$^{\rm r}$       &  $24350$            & $9.62$         & $6.70$        & $-$           & $7.60\pm0.28$ & $<7.14$      & $7.82\pm0.41$  & $<7.90$       & $7.20$         \\
EC\,20106$-$5248        &  $24500$            & $9.23$         & $7.37\pm0.31$ & $<8.42$       & $7.58\pm0.11$ & $-$          & $8.21\pm0.20$  & $-$           & $6.50$         \\
BD$+$48\,2721           &  $24800$            & $9.77$         & $6.50$        & $<8.48$       & $7.74\pm0.12$ & $-$          & $7.91\pm0.13$  & $-$           & $6.60$         \\
HD\,205805              &  $25000$            & $10.00$        & $8.22\pm0.35$ & $8.75\pm0.07$ & $8.18\pm0.20$ & $-$          & $7.49\pm0.20$  & $-$           & $6.40$         \\
PB\,7352$^{\rm r}$      &  $25000$            & $9.47$         & $6.95\pm0.07$ & $<8.33$       & $7.57\pm0.18$ & $-$          & $8.00\pm0.12$  & $-$           & $6.50$        \\
TON\,S\,135$^{\rm r}$   &  $25000$            & $9.40$         & $7.05\pm0.07$ & $-$           & $7.78\pm0.11$ & $-$          & $8.29\pm0.22$  & $-$           & $6.50$         \\
HE\,0321$-$0918         &  $25114$            & $8.98$         & $6.80$        & $-$           & $7.28\pm0.13$ & $-$          & $7.55\pm0.07$  & $7.75\pm0.07$ & $<5.95$        \\
LB\,1516$^{\rm r,l}$    &  $25200$            & $9.22$         & $6.70$        & $-$           & $7.89\pm0.18$ & $-$          & $7.83\pm0.16$  & $-$           & $6.40$         \\
PG\,0001$+$275          &  $25400$            & $9.10$         & $<7.51$       & $-$           & $7.38\pm0.16$ & $-$          & $7.68\pm0.17$  & $-$           & $6.10$         \\
PG\,1653$+$131          &  $25400$            & $9.30$         & $6.90$        & $<8.59$       & $7.86\pm0.25$ & $-$          & $8.06\pm0.34$  & $-$           & $6.90$        \\
HE\,0532$-$4503$^{\rm r}$ &  $25710$          & $8.93$         & $6.30$        & $<7.30$       & $7.06\pm0.18$ & $<6.74$      & $7.61\pm0.17$  & $7.15\pm0.07$ & $6.60$        \\
PG\,0342$+$026          &  $26000$            & $9.31$         & $7.20\pm0.26$ & $-$           & $7.97\pm0.19$ & $-$          & $7.76\pm0.21$  & $-$           & $6.60$        \\
GD\,108                 &  $26100$            & $8.54$         & $<5.96$       & $-$           & $6.65\pm0.07$ & $-$          & $<6.93$        & $-$           & $6.70$        \\
Feige\,65               &  $26200$            & $9.25$         & $7.00$        & $<8.70$       & $8.23\pm0.28$ & $-$          & $8.33\pm0.30$  & $-$           & $6.50$        \\
JL\,82$^{\rm r,l}$        &  $26500$            & $9.46$         & $6.90\pm0.14$ & $<8.35$       & $7.85\pm0.21$ & $-$          & $8.02\pm0.17$  & $-$           & $7.00$         \\
PHL\,457$^{\rm r,l}$    &  $26500$            & $9.46$         & $7.33\pm0.32$ & $-$           & $7.86\pm0.14$ & $-$          & $8.25\pm0.22$  & $-$           & $6.80$        \\
PG\,1248$+$164$^{\rm r}$ & $26600$            & $<8.00$        & $<7.98$       & $-$           & $8.01\pm0.21$ & $-$          & $7.90\pm0.22$  & $-$           & $6.50$        \\
PHL\,44$^{\rm l}$       &  $26600$            & $9.03$         & $<7.84$       & $-$           & $7.58\pm0.19$ & $-$          & $7.93\pm0.21$  & $-$           & $6.40$        \\
PG\,1432$+$159$^{\rm r}$ & $26900$            & $10.51$        & $<7.78$       & $-$           & $7.96\pm0.11$ & $-$          & $7.87\pm0.12$  & $-$           & $6.60$        \\
SB\,815                 &  $27000$            & $9.10$         & $6.50$        & $<7.29$       & $7.50\pm0.15$ & $-$          & $7.61\pm0.17$  & $-$           & $6.70$        \\
PG\,2205$+$023          &  $27100$            & $<8.00$        & $<6.90$       & $-$           & $7.10$        & $-$          & $7.20$         & $-$           & $6.70$        \\
PG\,2314$+$076          &  $27200$            & $<8.00$        & $<6.16$       & $-$           & $<6.79$       & $-$          & $<7.03$        & $-$           & $6.50$        \\
PG\,1716$+$426$^{\rm r,l}$ &  $27400$         & $9.48$         & $-$           & $<7.58$       & $7.73\pm0.45$ & $-$          & $8.00$         & $<8.33$       & $6.70$       \\
CPD$-$64$^{\rm \circ}$481$^{\rm r}$ & $27500$           & $9.50$         & $7.50\pm0.28$ & $8.20$        & $7.82\pm0.18$ & $8.25\pm0.07$ & $7.72\pm0.23$ & $<7.83$       & $6.40$       \\
PG\,0101$+$039$^{\rm r,l}$ & $27500$            & $9.34$         & $<6.56$       & $-$           & $7.50\pm0.21$ & $<8.35$      & $7.60\pm0.15$  & $<8.21$       & $6.40$       \\
PG\,2345$+$318$^{\rm r}$ & $27500$            & $9.78$         & $<7.78$       & $-$           & $7.39\pm0.23$ & $-$          & $7.80\pm0.28$  & $<8.22$       & $6.60$       \\
PG\,1743$+$477$^{\rm r}$ & $27600$            & $10.20$        & $<7.85$       & $-$           & $7.77\pm0.26$ & $<8.38$      & $7.94\pm0.18$  & $<8.23$       & $-$            \\
TON\,S\,183$^{\rm r}$   &  $27600$            & $9.20$         & $6.50$        & $-$           & $7.54\pm0.15$ & $-$          & $7.84\pm0.20$  & $<8.27$       & $6.60$         \\
HS\,2357$+$2201$^{\rm r}$ &  $27629$          & $9.46$         & $<6.33$       & $-$           & $7.55\pm0.13$ & $-$          & $7.76\pm0.17$  & $7.10$        & $6.30$       \\ 
EC\,14338$-$1445$^{\rm r}$ & $27700$          & $9.18$         & $7.00$        & $-$           & $7.62\pm0.17$ & $-$          & $7.51\pm0.25$  & $<8.61$       & $6.50$       \\
HD\,171858$^{\rm r}$    &  $27700$            & $9.20$         & $<6.11$       & $-$           & $7.53\pm0.12$ & $<8.51$      & $7.84\pm0.23$  & $<8.08$       & $6.80$       \\
SB\,485                 &  $27738$            & $9.50$         & $6.60$        & $-$           & $7.54\pm0.05$ & $<8.11$      & $7.64\pm0.17$  & $6.80$        & $6.40$       \\
EC\,03591$-$3232        &  $28000$            & $9.97$         & $7.80\pm0.32$ & $7.70$        & $8.10\pm0.16$ & $-$          & $7.78\pm0.29$  & $-$           & $7.30$        \\
\noalign{\smallskip}
\hline
\noalign{\smallskip}
Object              & $T_{\rm eff}$\,[K]      & He\,{\sc i/ii} & C\,{\sc ii}   & C\,{\sc iii}  & N\,{\sc ii}   & N\,{\sc iii} & O\,{\sc ii}    & Ne\,{\sc ii}  & Mg\,{\sc ii}  \\
\noalign{\smallskip}
\hline
\noalign{\smallskip}
EC\,12234$-$2607        &  $28000$            & $10.42$        & $7.70\pm0.26$ & $8.20$        & $7.78\pm0.18$ & $-$          & $8.52\pm0.23$  & $-$           & $7.00$        \\
PG\,2349$+$002          &  $28000$            & $8.55$         & $6.70$        & $<7.46$       & $7.10\pm0.26$ & $<8.25$      & $7.30$         & $-$           & $6.60$        \\
HE\,0136$-$2758         &  $28202$            & $<8.00$        & $-$           & $-$           & $7.05\pm0.26$ & $<8.73$      & $8.10\pm0.14$  & $<7.16$       & $<6.00$       \\
PG\,1549$-$001          &  $28252$            & $9.34$         & $7.20$        & $<7.60$       & $7.15\pm0.13$ & $<6.91$      & $7.93\pm0.16$  & $<7.30$       & $6.70$       \\
HE\,0016$+$0044         &  $28264$            & $9.34$         & $6.60$        & $<7.20$       & $6.95\pm0.13$ & $-$          & $8.04\pm0.22$  & $6.70$        & $6.60$       \\
HE\,2349$-$3135         &  $28520$            & $8.16$         & $<6.80$       & $-$           & $7.26\pm0.11$ & $<7.07$      & $7.61\pm0.17$  & $7.20$        & $6.60$       \\
PG\,1725$+$252$^{\rm r}$ & $28900$            & $9.00$         & $-$           & $<7.37$       & $7.45\pm0.51$ & $<8.32$      & $7.95\pm0.17$  & $<8.05$       & $6.70$       \\
HE\,0007$-$2212         &  $28964$            & $<8.00$        & $6.30$        & $-$           & $7.48\pm0.16$ & $<8.00$      & $7.51\pm0.19$  & $<6.80$       & $<6.11$       \\
PG\,1329$+$159$^{\rm r}$ & $29100$            & $9.60$         & $<7.90$       & $-$           & $7.42\pm0.60$ & $<8.04$      & $7.92\pm0.23$  & $<7.98$       & $-$  \\
LB\,275                 &  $29262$            & $9.54$         & $7.00$        & $7.30$        & $7.68\pm0.08$ & $<8.37$      & $7.60\pm0.14$  & $6.95\pm0.21$ & $6.40$ \\
Feige\,48$^{\rm r,s}$   &  $29500$            & $9.10$         & $7.50$        & $7.20$        & $7.28\pm0.12$ & $-$          & $7.65\pm0.20$  & $<7.94$       & $6.80$ \\
HE\,1421$-$1206$^{\rm r}$ & $29570$           & $<8.00$        & $6.80$        & $7.20$        & $7.15\pm0.24$ & $-$          & $7.47\pm0.16$  & $7.30\pm0.42$ & $6.60$ \\
PG\,0133$+$114$^{\rm r}$ & $29600$            & $9.70$         & $6.80$        & $<7.57$       & $7.80\pm0.20$ & $<8.27$      & $7.88\pm0.16$  & $-$           & $6.50$ \\
PG\,1101$+$249$^{\rm r}$ & $29600$            & $9.83$         & $-$           & $<7.31$       & $7.56\pm0.05$ & $<8.06$      & $7.77\pm0.26$  & $-$           & $6.60$ \\
HE\,0929$-$0424$^{\rm r}$ & $29602$       & $9.99$         & $6.60$        & $<7.53$       & $7.70\pm0.16$ & $<8.31$      & $7.64\pm0.14$  & $7.00$        & $-$ \\
PHL\,861$^{\rm r}$       & $29668$            & $<8.00$        & $<6.20$       & $-$           & $7.08\pm0.16$ & $-$          & $7.63\pm0.20$  & $<6.80$       & $6.70$ \\
PG\,1303$+$097          &  $29800$            & $9.83$         & $7.30$        & $<7.73$       & $7.98\pm0.17$ & $<8.32$      & $7.91\pm0.21$  & $<8.25$       & $7.20$ \\
HE\,2150$-$0238$^{\rm r}$ & $29846$       & $9.64$         & $<6.80$       & $-$           & $7.67\pm0.18$ & $<8.00$      & $7.10\pm0.17$  & $<7.20$       & $<6.32$ \\
PG\,1512$+$244$^{\rm r}$ &  $29900$           & $9.95$         & $8.20\pm0.14$ & $8.00$        & $8.02\pm0.12$ & $-$          & $7.84\pm0.13$  & $-$           & $7.00$ \\
HE\,2135$-$3749$^{\rm r}$ & $29924$           & $9.55$         & $<5.98$       & $-$           & $7.50\pm0.09$ & $7.90$       & $6.75\pm0.07$  & $6.50$        & $<5.79$ \\ 
HE\,2222$-$3738         &  $30248$            & $8.35$         & $7.10$        & $7.40$        & $7.73\pm0.08$ & $<8.18$      & $7.74\pm0.26$  & $<7.03$       & $7.20$ \\
HE\,1047$-$0436$^{\rm r}$ & $30280$       & $9.65$         & $7.30$        & $7.88$        & $7.92\pm0.12$ & $8.00\pm0.01$ & $7.70\pm0.13$ & $6.80$        & $6.90$ \\
PG\,1519$+$640$^{\rm r}$ &  $30300$           & $9.63$         & $6.60$        & $-$           & $7.46\pm0.14$ & $-$          & $7.89\pm0.16$  & $-$           & $6.50$ \\
HE\,2238$-$1455         &  $30393$            & $9.63$         & $<6.84$       & $-$           & $<6.79$       & $-$          & $6.95\pm0.13$  & $<6.95$       & $6.50$ \\ 
Feige\,38               &  $30600$            & $9.63$         & $7.73\pm0.35$ & $7.80$        & $8.02\pm0.14$ & $<8.29$      & $7.65\pm0.11$  & $<8.02$       & $7.10$ \\
PG\,1710$+$490          &  $30600$            & $9.57$         & $7.20$        & $<6.99$        & $7.94\pm0.15$ & $8.20$       & $7.80\pm0.09$  & $-$           & $6.90$ \\
EC\,14248$-$2647        &  $31400$            & $10.36$        & $<6.61$       & $-$           & $7.48\pm0.17$ & $7.75\pm0.07$ & $<6.76$       & $<7.81$       & $<6.10$ \\
HE\,0230$-$4323$^{\rm r,s}$   & $31552$       & $9.42$         & $6.70$        & $<6.71$       & $7.28\pm0.13$ & $7.55\pm0.07$ & $7.51\pm0.18$ & $6.90\pm0.14$ & $6.90$ \\
KPD\,2109$+$440$^{\rm s}$     & $31800$       & $9.78$         & $<6.74$       & $-$           & $6.95\pm0.21$ & $<8.92$      & $<6.83$        & $<7.86$       & $<6.20$ \\
$[$CW83$]$\,1758$+$36         & $32000$       & $10.49$        & $<8.50$       & $7.70$        & $7.66\pm0.07$ & $-$          & $<6.91$        & $-$           & $-$ \\
EC\,21043$-$4017        &  $32400$            & $10.42$        & $-$           & $<6.76$       & $7.70\pm0.14$ & $7.85\pm0.07$ & $7.10$        & $<7.88$       & $<6.80$ \\
EC\,20229$-$3716        &  $32500$            & $10.25$        & $8.15\pm0.30$ & $8.60\pm0.16$ & $8.07\pm0.19$ & $8.27\pm0.06$ & $7.62\pm0.15$ & $-$           & $6.80$ \\
PG\,1116$+$301$^{\rm r}$ &  $32500$           & $9.78$         & $<8.05$       & $7.30$        & $7.83\pm0.46$ & $7.90$       & $<6.91$        & $-$           & $<6.53$ \\
PG\,2151$+$100          &  $32700$            & $<9.00$        & $<6.80$       & $7.25\pm0.35$ & $<6.92$       & $-$          & $7.36\pm0.15$  & $<7.83$       & $<6.34$ \\ 
HS\,2033$+$0821         &  $32706$            & $10.44$        & $7.60$        & $8.80\pm0.26$ & $7.63\pm0.20$ & $7.80\pm0.14$ & $<6.93$       & $<6.70$       & $<6.87$ \\
EC\,05479$-$5818        &  $33000$            & $10.34$        & $7.90\pm0.58$ & $8.30\pm0.10^{\rm p}$ & $7.94\pm0.18$ & $7.75\pm0.07$ & $<6.90$       & $<7.85$       & $<6.47$ \\
PHL\,932                &  $33644$            & $10.36$        & $-$           & $<6.34$       & $7.37\pm0.10$ & $7.90$       & $6.70$         & $4.00$        & $<6.16$ \\
EGB\,5$^{\rm r}$        &  $34060$            & $9.23$         & $-$           & $<6.40$       & $7.07\pm0.15$ & $7.73\pm0.23$ & $7.85\pm0.35$ & $<7.23$       & $<6.14$ \\
PHL\,555                &  $34126$            & $10.64$        & $<6.50$       & $6.90$        & $7.70\pm0.06$ & $7.65\pm0.07$ & $7.05\pm0.07$ & $6.70$        & $<6.57$ \\
PG\,1219$+$534$^{\rm s}$ &  $34300$           & $10.40$        & $-$           & $<6.30$       & $7.50\pm0.08$ & $7.80$       & $7.45\pm0.21$  & $<7.45$       & $<6.55$ \\
HE\,1050$-$0630         &  $34501$            & $10.60$        & $<6.88$       & $7.10$        & $7.56\pm0.18$ & $7.70\pm0.14$ & $<6.85$       & $<6.80$       & $<6.61$ \\
HE\,1450$-$0957         &  $34563$            & $10.71$        & $-$           & $<6.96$       & $7.62\pm0.10$ & $7.70\pm0.42$ & $<7.19$       & $<7.08$       & $<6.96$ \\
EC\,13047$-$3049        &  $34700$            & $9.43$         & $<7.79$       & $6.30$        & $<7.07$       & $<7.07$      & $<7.05$        & $<7.83$       & $6.80$ \\
HE\,1448$-$0510$^{\rm r}$  &  $34760$         & $8.59$         & $<7.04$       & $6.10$        & $7.50\pm0.28$ & $7.20$       & $<6.98$        & $7.15\pm0.35$ & $7.00$ \\ 
PHL\,334                &  $34800$            & $10.58$        & $-$           & $<6.29$       & $7.52\pm0.08$ & $7.65\pm0.07$ & $<6.92$       & $<7.46$       & $<6.75$ \\
\noalign{\smallskip}
\hline
\noalign{\smallskip}
Object              & $T_{\rm eff}$\,[K]      & He\,{\sc i/ii} & C\,{\sc ii}   & C\,{\sc iii}  & N\,{\sc ii}   & N\,{\sc iii} & O\,{\sc ii}    & Ne\,{\sc ii}  & Mg\,{\sc ii}  \\
\noalign{\smallskip}
\hline
\noalign{\smallskip}
HE\,2151$-$1001         &  $34984$            & $10.40$        & $6.90$        & $<6.68$       & $7.40\pm0.14$ & $8.00\pm0.14$ & $6.70$        & $<6.97$       & $<6.93$  \\
Feige\,49               &  $35000$            & $11.18$        & $7.60$        & $8.20\pm0.17$ & $8.22\pm0.19$ & $-$           & $8.58\pm0.16$  & $8.00$        & $7.50$  \\
PG\,0909$+$164          &  $35300$            & $9.24$         & $-$           & $<6.47$       & $<7.04$       & $-$          & $7.90\pm0.28$  & $<7.83$       & $7.00$ \\
CD$-$24$^{\rm \circ}$731$^{\rm r}$ & $35400$  & $9.10$         & $<7.14$       & $<6.15$       & $<7.60$       & $7.20\pm0.10$ & $<7.51$       & $<7.41$       & $<6.54$  \\
HE\,1021$-$0255         &  $35494$            & $10.54$        & $-$           & $<6.70$       & $<7.20$       & $7.40$       & $<7.12$        & $<7.00$       & $<6.98$ \\
HD\,149382              &  $35500$            & $10.56$        & $<7.13$       & $6.60$        & $7.67\pm0.14$ & $7.60\pm0.17$ & $7.55\pm0.21$ & $<7.11$       & $<6.17$ \\
PG\,0909$+$276          &  $35500$            & $11.00$        & $8.45\pm0.07$ & $8.93\pm0.13$ & $7.83\pm0.13$ & $7.65\pm0.07$ & $7.50$        & $<7.87$       & $-$ \\
HE\,0101$-$2707         &  $35568$            & $11.08$        & $8.30$        & $9.23\pm0.21^{\rm p}$ & $7.42\pm0.13$ & $7.65\pm0.07$ & $<7.20$       & $<6.60$       & $<6.86$ \\
HE\,0019$-$5545         &  $35662$            & $10.56$        & $-$           & $8.05\pm0.07$ & $7.40$        & $7.40$       & $<7.19$        & $7.10$        & $<6.97$ \\
PG\,1207$-$032          &  $35693$            & $10.52$        & $<7.01$       & $7.10$        & $7.25\pm0.13$ & $7.35\pm0.07$ & $7.25\pm0.35$ & $<6.90$       & $<6.79$ \\
GD\,619                 &  $36097$            & $10.67$        & $<7.00$       & $7.20$        & $7.15\pm0.21$ & $7.15\pm0.07$ & $7.20$        & $6.70$        & $<6.77$ \\
HE\,0123$-$3330         &  $36602$            & $10.51$        & $-$           & $7.80$        & $7.50$        & $7.30\pm0.28$ & $7.55\pm0.35$ & $<6.93$       & $<6.88$ \\
PG\,1505$+$074          &  $37100$            & $9.31$         & $-$           & $<6.11$       & $<6.86$       & $-$           & $<6.78$       & $-$           & $7.00$ \\
HE\,1407$+$0033         &  $37309$            & $9.01$         & $-$           & $<6.49$       & $<7.07$       & $-$           & $<7.28$       & $<7.00$       & $<7.02$ \\
PHL\,1548               &  $37400$            & $10.45$        & $7.50$        & $7.50\pm0.10$ & $7.71\pm0.25$ & $7.20\pm0.14$ & $<7.25$       & $<7.81$       & $<6.83$ \\
$[$CW83$]$\,0512$-$08   &  $38400$            & $11.27$        & $9.00\pm0.14$ & $8.50\pm0.10$ & $8.05\pm0.16$ & $7.50\pm0.30$ & $<7.32$       & $-$           & $-$ \\
PB\,5333                &  $40600$            & $9.38$         & $-$           & $<6.44$       & $<7.57$       & $7.05\pm0.21$ & $<7.51$       & $<7.89$       & $-$ \\
\noalign{\smallskip}
\hline
\end{longtable}
\tablefoot{$\dag$Spectra with low S/N. $^{\rm r}$Radial velocity variable sdB. $^{\rm s}$Short-period pulsator of sdBV$_{\rm r}$-type. $^{\rm l}$Long-period pulsator of sdBV$_{\rm s}$-type. $^{\rm p}$Peculiar line profiles due to stratification.}

\newpage

\begin{longtable}{llllllllll}
\hline
\noalign{\smallskip}
Object                   & Al\,{\sc iii} & Si\,{\sc iii} & Si\,{\sc iv} & P\,{\sc iii} & S\,{\sc ii}   & S\,{\sc iii}  & Ar\,{\sc ii}  & K\,{\sc ii} & Ca\,{\sc iii} \\
\noalign{\smallskip}
\hline
\noalign{\smallskip}
HE\,0151$-$3919          & $<4.74$       & $<6.04$       & $-$          & $<5.09$      & $6.40$        & $6.90$        & $6.73\pm0.21$ & $<5.96$    & $-$            \\
EC\,21494$-$7018         & $<6.48$       & $6.20$        & $-$          & $<5.87$      & $<6.37$       & $-$           & $<6.85$       & $<6.94$    & $-$            \\
EC\,15103$-$1557         & $<5.90$       & $6.05\pm0.21$ & $-$          & $<5.25$      & $<5.79$       & $-$           & $<6.00$       & $<6.06$    & $-$            \\
EC\,11349$-$2753         & $5.80$        & $6.97\pm0.23$ & $-$          & $<4.82$      & $6.75\pm0.24$ & $6.80\pm0.14$ & $6.80\pm0.10$ & $5.60$     & $-$            \\
HD\,4539                 & $5.60$        & $6.70\pm0.08$ & $<6.50$      & $<4.82$      & $7.10\pm0.14$ & $6.55\pm0.07$ & $7.17\pm0.15$ & $5.80$     & $-$            \\
HE\,2307$-$0340          & $5.35\pm0.07$ & $<6.21$       & $-$          & $5.90$       & $<6.39$       & $-$           & $6.30$        & $6.50$     & $-$            \\
HE\,0539$-$4246          & $<4.94$       & $<6.32$       & $-$          & $5.90$       & $<6.38$       & $-$           & $6.80$        & $<6.40$    & $-$            \\
EC\,14345$-$1729         & $<5.86$       & $7.28\pm0.15$ & $-$          & $<5.76$      & $6.75\pm0.21$ & $6.95\pm0.07$ & $6.53\pm0.06$ & $<5.85$    & $-$            \\
PG\,1627$+$017$^{\rm r,p}$ & $5.50$        & $6.87\pm0.06$ & $-$          & $-$          & $6.50$        & $-$           & $-$           & $-$        & $-$            \\
PG\,1432$+$004           & $6.00$        & $7.25\pm0.13$ & $<7.13$      & $<4.84$      & $6.83\pm0.10$ & $6.80\pm0.14$ & $6.90\pm0.17$ & $<5.71$    & $-$            \\
HE\,2208$+$0126$^{\rm r}$          & $5.00$        & $<6.17$       & $-$          & $<5.47$      & $<6.43$       & $-$           & $<6.26$       & $<6.10$    & $-$            \\
GD\,687$^{\rm r}$        & $<5.10$       & $<6.11$       & $-$          & $<5.88$      & $<6.75$       & $-$           & $<6.95$       & $<6.19$    & $-$            \\
EC\,20106$-$5248         & $5.60$        & $6.96\pm0.05$ & $<6.95$      & $<5.13$      & $6.30$        & $6.55\pm0.07$ & $<6.28$       & $<5.86$    & $-$            \\
BD$+$48\,2721            & $5.40$        & $7.13\pm0.21$ & $-$          & $-$          & $6.03\pm0.15$ & $<5.73$       & $6.77\pm0.23$ & $<5.57$    & $-$            \\
HD\,205805               & $5.60\pm0.14$ & $6.97\pm0.30^{\rm p}$ & $<6.50$      & $<4.74$      & $6.68\pm0.22$ & $6.85\pm0.07$ & $6.73\pm0.12$ & $<5.06$    & $-$            \\
PB\,7352$^{\rm r}$       & $5.50$        & $6.85\pm0.06$ & $<7.07$      & $<5.04$      & $6.75\pm0.49$ & $6.45\pm0.21$ & $6.40$        & $<5.82$    & $-$            \\
TON\,S\,135$^{\rm r}$    & $5.80$        & $7.17\pm0.15$ & $-$          & $<5.33$      & $<6.45$       & $<6.45$       & $-$           & $<6.20$    & $-$            \\
HE\,0321$-$0918          & $<4.98$       & $6.00$        & $-$          & $<5.20$      & $<6.45$       & $-$           & $<6.60$       & $<6.11$    & $-$            \\ 
LB\,1516$^{\rm r,l}$     & $<6.38$       & $6.87\pm0.06$ & $-$          & $<5.59$      & $<6.84$       & $6.40$        & $<6.88$       & $<6.12$    & $-$            \\
PG\,0001$+$275           & $<5.76$       & $6.45\pm0.07$ & $-$          & $<5.42$      & $<6.43$       & $6.00$        & $<6.48$       & $<6.35$    & $-$            \\ 
PG\,1653$+$131$^{\rm r}$ & $5.50$        & $6.95\pm0.10$ & $<7.41$      & $<5.89$      & $-$           & $<6.51$       & $<6.88$       & $<6.07$    & $-$            \\
HE\,0532$-$4503$^{\rm r}$ & $<4.17$      & $<5.40$       & $<5.40$      & $<4.81$      & $<6.27$       & $5.92\pm0.07$ & $<5.98$       & $<5.69$    & $-$            \\
PG\,0342$+$026           & $<5.67$       & $6.99\pm0.18$ & $<6.85$      & $<4.81$      & $6.83\pm0.19$ & $6.70$        & $6.80\pm0.10$ & $<5.61$    & $-$            \\
GD\,108                  & $<5.78$       & $5.95\pm0.07$ & $<6.99$      & $<4.99$      & $-$           & $<6.26$       & $<6.24$       & $<5.79$    & $-$            \\
Feige\,65                & $<5.88$       & $7.18\pm0.17$ & $<7.37$      & $<5.19$      & $6.65\pm0.07$ & $<6.49$       & $6.85\pm0.07$ & $<6.11$    & $-$            \\
JL\,82$^{\rm r,l}$         & $<5.84$       & $6.98\pm0.17$ & $<7.26$      & $<5.54$      & $<6.55$       & $6.20$        & $<6.57$       & $<5.90$    & $-$            \\
PHL\,457$^{\rm r}$       & $5.70$        & $6.97\pm0.17$ & $<7.07$      & $<5.02$      & $<6.35$       & $6.30$        & $<6.74$       & $<5.82$    & $-$            \\
PG\,1248$+$164$^{\rm r}$ & $5.80$        & $6.88\pm0.13$ & $-$          & $-$          & $7.00$        & $6.50$        & $6.90$        & $-$        & $-$            \\
PHL\,44$^{\rm l}$        & $<5.81$       & $6.73\pm0.12$ & $-$          & $<5.04$      & $<6.45$       & $6.30$        & $<6.34$       & $<5.86$    & $-$            \\
PG\,1432$+$159$^{\rm r}$ & $6.00$        & $6.87\pm0.15$ & $-$          & $-$          & $<6.84$       & $6.50$        & $7.00$        & $-$         & $-$           \\
SB\,815                  & $<5.79$       & $<5.54$       & $<5.94$      & $<4.94$      & $<6.89$       & $5.95\pm0.07$ & $<6.78$       & $<5.90$    & $-$            \\
PG\,2205$+$023           & $<6.11$       & $6.93\pm0.10$ & $<7.31$      & $<5.34$      & $-$           & $<5.95$       & $<6.94$       & $<6.24$    & $-$           \\
PG\,2314$+$076           & $5.70$        & $6.45\pm0.17$ & $-$          & $<5.06$      & $<6.52$       & $-$           & $<6.57$       & $6.40$     & $-$            \\
PG\,1716$+$426$^{\rm r,l}$ &  $-$        & $6.40$        & $-$          & $-$          & $-$           & $<6.01$       & $-$           & $-$        & $-$           \\
CPD$-$64$^{\rm \circ}$481$^{\rm r}$ & $5.40\pm0.14$ & $6.83\pm0.28^{\rm p}$ & $6.40$ & $<4.75$ & $6.58\pm0.36$ & $6.65\pm0.07$ & $6.50\pm0.10$ & $<5.11$    & $<8.63$       \\
PG\,0101$+$039$^{\rm r}$ & $<5.75$       & $6.40\pm0.01$ & $-$          & $<5.00$      & $<6.63$       & $-$           & $<6.76$       & $-$        & $-$           \\
PG\,2345$+$318$^{\rm r}$ & $-$           & $6.77\pm0.10$ & $-$          & $-$          & $<6.74$       & $-$           & $<6.71$       & $-$        & $-$           \\ 
PG\,1743$+$477$^{\rm r}$ & $5.80$        & $6.83\pm0.15^{\rm p}$ & $-$          & $-$          & $<6.38$       & $-$           & $<6.26$       & $-$        & $-$            \\
TON\,S\,183$^{\rm r}$    & $5.40$        & $6.60$        & $-$          & $<4.97$      & $<6.59$       & $6.05\pm0.07$ & $<6.58$       & $<5.84$    & $-$            \\
HS\,2357$+$2201$^{\rm r}$ & $<4.87$      & $<5.86$       & $-$          & $<5.10$      & $<6.11$       & $6.50$        & $-$           & $<5.96$    & $-$           \\
EC\,14338$-$1445$^{\rm r}$ & $<5.96$     & $7.37\pm0.42$ & $-$          & $<5.27$      & $6.80$        & $6.75\pm0.07$ & $<6.95$       & $<6.08$    & $-$           \\ 
HD\,171858$^{\rm r}$     & $5.30$        & $6.72\pm0.10$ & $<6.05$      & $<4.76$      & $<6.28$       & $6.05\pm0.07$ & $<6.01$       & $<5.69$    & $-$           \\
SB\,485                  & $<4.70$       & $<5.46$       & $-$          & $<4.82$      & $<6.48$       & $6.10\pm0.28$ & $6.70$        & $<5.74$    & $-$           \\
EC\,03591$-$3232         & $6.40\pm0.14$ & $7.91\pm0.38^{\rm p}$ & $7.00\pm0.28$ & $<4.83$     & $7.18\pm0.08$ & $6.90\pm0.14$ & $7.03\pm0.15$ & $<5.76$    & $-$          \\
\noalign{\smallskip}
\hline
\noalign{\smallskip}
Object                   & Al\,{\sc iii} & Si\,{\sc iii} & Si\,{\sc iv} & P\,{\sc iii} & S\,{\sc ii}   & S\,{\sc iii}  & Ar\,{\sc ii}  & K\,{\sc ii} & Ca\,{\sc iii} \\
\noalign{\smallskip}
\hline
\noalign{\smallskip}
EC\,12234$-$2607         & $<6.09$       & $7.29\pm0.35$ & $<5.92$      & $<5.47$      & $7.03\pm0.05$ & $6.90\pm0.14$ & $<7.01$       & $<6.23$    & $-$          \\
PG\,2349$+$002           & $<5.92$       & $6.47\pm0.21$ & $<6.14$      & $<5.15$      & $-$           & $<5.98$       & $<6.85$       & $<6.04$    & $-$           \\
HE\,0136$-$2758          & $<4.89$       & $6.40\pm0.14$ & $<6.42$      & $<5.05$      & $<6.90$       & $6.30\pm0.14$ & $<6.82$       & $<6.06$    & $-$          \\
PG\,1549$-$001           & $5.30\pm0.14$ & $6.67\pm0.49$ & $6.90\pm0.14$ & $<5.07$     & $7.10$        & $6.90\pm0.14$ & $7.00\pm0.14$ & $6.45\pm0.21$ & $-$       \\
HE\,0016$+$0044          & $<4.90$       & $5.60$        & $<6.25$      & $4.90$       & $-$           & $6.00$        & $<6.85$       & $6.25\pm0.07$ & $-$       \\
HE\,2349$-$3135          & $<5.14$       & $<5.95$       & $-$          & $<5.40$      & $-$           & $<5.87$       & $<7.08$       & $6.00$     & $-$          \\
PG\,1725$+$252$^{\rm r}$ & $5.60$        & $6.47\pm0.06$ & $-$          & $-$          & $7.40$        & $-$           & $<6.81$       & $-$        & $-$          \\
HE\,0007$-$2212          & $<5.10$       & $6.70\pm0.28$ & $7.25\pm0.21$ & $<5.21$     & $<7.29$       & $7.05\pm0.07$ & $6.85\pm0.07$ & $<5.80$    & $-$          \\
PG\,1329$+$159$^{\rm r}$ & $<5.88$       & $6.53\pm0.15$ & $-$          & $-$          & $<5.78$       & $<5.78$       & $<6.80$       & $-$        & $-$          \\
LB\,275                  & $<4.98$       & $7.07\pm0.23$ & $6.60$       & $<5.15$      & $7.30$        & $6.65\pm0.07$ & $6.90$        & $<6.06$    & $-$          \\ 
Feige\,48$^{\rm r}$      & $5.60$        & $6.55\pm0.06$ & $6.20$       & $<4.98$      & $<7.23$       & $5.95\pm0.07$ & $<6.82$       & $<5.93$    & $-$ \\
HE\,1421$-$1206$^{\rm r}$ & $5.50$       & $6.20$        & $<6.75$      & $<5.10$      & $-$           & $5.90$        & $<7.00$       & $6.60$     & $<8.50$ \\
PG\,0133$+$114$^{\rm r}$ & $5.50$        & $7.50\pm0.37^{\rm p}$ & $7.30$       & $<4.76$      & $6.90\pm0.14$ & $6.95\pm0.07$ & $6.70\pm0.10$ & $<5.51$    & $-$ \\ 
PG\,1101$+$249$^{\rm r}$ & $5.60$        & $7.07\pm0.06$ & $<7.60$      & $-$          & $<6.92$       & $6.50$        & $6.80$        & $-$        & $-$ \\ 
HE\,0929$-$0424$^{\rm r}$ & $5.50$   & $<6.30$       & $6.20\pm0.28$ & $<5.17$     & $<7.30$       & $6.45\pm0.07$ & $7.15\pm0.21$ & $6.40$     & $8.70$ \\
PHL\,861$^{\rm r}$       & $5.50$        & $<5.60$       & $<6.21$      & $<5.00$      & $-$           & $6.00$        & $<7.08$       & $<6.28$    & $-$ \\
PG\,1303$+$097           & $5.90$        & $7.40\pm0.39^{\rm p}$ & $<8.13$      & $<5.51$      & $<7.43$       & $6.85\pm0.07$ & $<7.12$       & $<6.45$    & $-$ \\
HE\,2150$-$0238$^{\rm r}$ & $<5.30$  & $6.40\pm0.28$ & $<6.40$      & $<5.50$      & $<7.40$       & $6.85\pm0.21$ & $7.78\pm0.02$ & $6.40$     & $<8.40$ \\
PG\,1512$+$244$^{\rm r}$ & $5.80$        & $7.56\pm0.22$ & $-$          & $-$          & $7.20$        & $6.90$        & $7.10$        & $-$        & $-$ \\
HE\,2135$-$3749$^{\rm r}$ & $5.10$   & $<5.46$       & $<5.78$      & $<4.80$      & $7.00$        & $6.70$        & $7.33\pm0.29$ & $6.30\pm0.10$ & $-$ \\
HE\,2222$-$3738          & $<5.20$       & $6.73\pm0.15$ & $6.70\pm0.42$ & $5.50$      & $7.00$        & $6.80\pm0.14$ & $7.10\pm0.17$ & $<6.27$    & $-$ \\
HE\,1047$-$0436$^{\rm r}$ & $5.40$   & $6.83\pm0.25$ & $6.77\pm0.38$ & $<5.02$     & $7.60$        & $7.15\pm0.07$ & $7.04\pm0.16$ & $<6.00$    & $-$ \\
PG\,1519$+$640$^{\rm r}$ & $<5.80$       & $6.93\pm0.15$ & $-$          & $<5.38$      & $<6.70$       & $6.80\pm0.14$ & $<6.91$       & $<5.88$    & $-$ \\
HE\,2238$-$1455          & $<5.06$       & $<5.89$       & $5.70$       & $<5.08$      & $-$           & $5.70$        & $<6.91$       & $6.10$     & $-$ \\
Feige\,38                & $5.80$        & $7.80\pm0.34^{\rm p}$ & $7.15\pm0.07$ & $<4.91$     & $7.26\pm0.15$ & $7.00$        & $7.10\pm0.17$ & $<5.82$    & $-$ \\
PG\,1710$+$490           & $6.30$        & $7.60\pm0.20^{\rm p}$ & $-$           & $-$         & $<7.48$       & $6.80$        & $<7.06$       & $<6.33$    & $-$ \\
EC\,14248$-$2647         & $<5.87$       & $5.70$        & $<5.89$       & $-$         & $<7.31$       & $6.80$        & $7.47\pm0.15$ & $6.20$     & $8.30$ \\
HE\,0230$-$4323$^{\rm r,s}$   & $<5.20$  & $6.20\pm0.20$ & $6.60\pm0.44$ & $<5.24$     & $-$           & $5.90\pm0.14$ & $<7.07$       & $<6.26$    & $<7.90$ \\
KPD\,2109$+$440$^{\rm s}$     & $<5.99$  & $<5.75$       & $-$           & $<5.37$     & $<7.55$       & $6.15\pm0.07$ & $<7.14$       & $<6.38$    & $<8.38$ \\
$[$CW83$]$\,1758$+$36         & $<6.14$  & $-$           & $<5.94$       & $<5.60$     & $<7.54$       & $7.05\pm0.07$ & $8.10\pm0.36$ & $6.95\pm0.21$ & $-$ \\
EC\,21043$-$4017         & $<6.23$       & $<5.85$       & $-$           & $<5.77$     & $<7.92$       & $6.95\pm0.07$ & $7.90\pm0.14$ & $<6.95$    & $8.10$ \\
EC\,20229$-$3716         & $6.00$        & $7.00\pm0.33$ & $7.10\pm0.10$ & $<4.96$     & $<7.42$       & $6.90$        & $7.00$        & $<6.16$    & $<7.33$ \\
PG\,1116$+$301$^{\rm r}$ & $<6.53$       & $6.85\pm0.07$ & $<7.18$       & $-$         & $<7.84$       & $-$           & $<7.38$       & $-$        & $-$ \\
PG\,2151$+$100           & $<6.05$       & $6.10$        & $<5.91$       & $<5.44$     & $-$           & $<5.51$       & $<7.29$       & $<6.60$    & $<8.22$ \\
HS\,2033$+$0821          & $5.60$        & $<6.10$       & $5.95\pm0.35$ & $<5.59$     & $7.90$        & $7.60\pm0.01$ & $8.03\pm0.21$ & $7.20$     & $-$ \\
EC\,05479$-$5818         & $<6.51$       & $6.00$        & $<5.94$       & $<5.52$     & $8.10\pm0.16$ & $7.40^{\rm p}$  & $8.03\pm0.15$ & $6.95\pm0.35$ & $<8.17$ \\
PHL\,932                 & $<5.39$       & $<6.35$       & $-$           & $<5.40$     & $-$           & $6.80$        & $7.15\pm0.07$ & $<6.53$    & $8.20\pm0.14$ \\ 
EGB\,5$^{\rm r}$         & $<6.16$       & $-$           & $<5.39$       & $<5.38$     & $<8.18$       & $6.15\pm0.07$ & $<7.25$       & $<6.54$    & $7.95\pm0.07$ \\
PHL\,555                 & $<5.74$       & $-$           & $<5.34$       & $6.10$      & $-$           & $7.30\pm0.14$ & $7.93\pm0.21$ & $<6.81$    & $8.10\pm0.42$ \\
PG\,1219$+$534$^{\rm s}$ & $<6.19$       & $-$           & $<5.06$       & $5.80$      & $-$           & $6.40\pm0.14$ & $<7.77$       & $-$        & $7.65\pm0.07$ \\
HE\,1050$-$0630          & $5.80$        & $<6.09$       & $5.45\pm0.07$ & $5.80$      & $-$           & $7.20\pm0.01$ & $7.55\pm0.07$ & $7.30$     & $8.10$ \\
HE\,1450$-$0957          & $5.90\pm0.14$ & $<6.85$       & $6.50\pm0.28$ & $<6.04$     & $-$           & $6.00$        & $7.65\pm0.07$ & $<6.50$    & $<8.34$ \\
EC\,13047$-$3049         & $6.30$        & $<5.98$       & $6.10\pm0.14$ & $<5.84$     & $-$           & $<6.32$       & $<7.90$       & $<7.23$    & $<7.98$ \\
HE\,1448$-$0510$^{\rm r}$  & $<5.80$     & $6.65\pm0.07$ & $6.40\pm0.10$ & $<5.87$     & $-$           & $<5.89$       & $<7.93$       & $7.15\pm0.07$ & $7.40$ \\
PHL\,334                 & $<6.24$       & $-$           & $<5.35$       & $<5.75$     & $-$           & $7.20$        & $<7.76$       & $<6.88$    & $7.80\pm0.28$ \\
\noalign{\smallskip}
\hline
\noalign{\smallskip}
Object                   & Al\,{\sc iii} & Si\,{\sc iii} & Si\,{\sc iv} & P\,{\sc iii} & S\,{\sc ii}   & S\,{\sc iii}  & Ar\,{\sc ii}  & K\,{\sc ii} & Ca\,{\sc iii} \\
\noalign{\smallskip}
\hline
\noalign{\smallskip}
HE\,2151$-$1001          & $5.85\pm0.21$ & $-$           & $<5.57$       & $<5.97$     & $-$           & $6.60$        & $<7.60$       & $7.50$     & $<8.12$ \\
Feige\,49                & $<6.80$       & $7.17\pm0.13$ & $<6.78$       & $<5.78$     & $-$           & $<7.00$       & $<7.85$       & $<6.93$    & $<7.41$ \\
PG\,0909$+$164           & $<6.47$       & $<5.99$       & $6.40\pm0.14$ & $<5.83$     & $-$           & $<5.80$       & $<7.85$       & $<7.00$    & $<7.41$ \\
CD$-$24$^{\rm \circ}$731$^{\rm r}$ & $<6.28$ & $-$       & $<4.91$       & $<5.91$     & $-$           & $6.60\pm0.01$ & $<8.47$       & $<7.55$     & $7.55\pm0.07$ \\
HE\,1021$-$0255          & $<5.93$       & $-$           & $<5.47$       & $<5.94$     & $-$           & $7.28$        & $8.00$        & $<7.13$    & $8.00$ \\
HD\,149382               & $<5.96$       & $<5.12$       & $5.40$        & $-$         & $-$           & $7.35\pm0.07$ & $7.70$        & $6.50$     & $7.90\pm0.14$ \\
PG\,0909$+$276           & $<6.82$       & $6.90\pm0.14$ & $-$           & $<6.15$     & $-$           & $7.65\pm0.07$ & $8.20$        & $<7.50$    & $8.15\pm0.07$ \\
HE\,0101$-$2707          & $<5.92$       & $<6.20$       & $6.05\pm0.07$ & $6.20$      & $-$           & $7.75\pm0.07^{\rm p}$ & $7.97\pm0.12$ & $7.55\pm0.07$ & $8.30\pm0.28$ \\
HE\,0019$-$5545          & $<6.00$       & $<6.85$       & $6.10\pm0.14$ & $6.10$      & $-$           & $6.70$        & $<7.60$       & $<7.22$    & $7.90\pm0.28$ \\
PG\,1207$-$032           & $<5.82$       & $<6.35$       & $5.70$        & $<5.84$     & $-$           & $7.50\pm0.01$ & $<7.90$       & $7.00\pm0.28$ & $7.75\pm0.35$ \\
GD\,619                  & $<5.80$       & $<5.55$       & $5.20$        & $6.20$      & $-$           & $7.40\pm0.01$ & $<7.87$       & $<6.97$    & $8.25\pm0.07$ \\
HE\,0123$-$3330          & $6.20$        & $<6.80$       & $6.10\pm0.30$ & $<6.01$     & $-$           & $6.20$        & $<8.14$       & $7.20\pm0.57$ & $7.55\pm0.07$ \\  
PG\,1505$+$074           & $<5.91$       & $<5.98$       & $6.25\pm0.07$ & $-$         & $-$           & $<5.47$       & $-$           & $<6.94$    & $-$ \\
HE\,1407$+$0033          & $<6.04$       & $-$           & $<6.10$       & $6.40$      & $-$           & $<6.39$       & $<8.45$       & $<7.88$    & $<7.47$ \\
PHL\,1548                & $<6.77$       & $<6.80$       & $5.80$        & $<6.02$     & $-$           & $7.90\pm0.14$ & $<8.86$       & $7.50$     & $8.15\pm0.07$ \\
$[$CW83$]$\,0512$-$08    & $<6.86$       & $<7.48$       & $5.50$        & $<6.09$     & $-$           & $8.05\pm0.21$ & $-$           & $-$        & $8.15\pm0.07$ \\
PB\,5333                 & $<6.85$       & $-$           & $<5.26$       & $<6.16$     & $-$           & $<6.37$       & $<8.60$       & $<7.85$    & $<7.37$ \\ 
\hline
\end{longtable}
\tablefoot{$\dag$Spectra with low S/N. $^{\rm r}$Radial velocity variable sdB. $^{\rm s}$Short-period pulsator of sdBV$_{\rm r}$-type. $^{\rm l}$Long-period pulsator of sdBV$_{\rm s}$-type.}

\newpage

\begin{longtable}{llllllllll}
\hline
\noalign{\smallskip}
Object             & Sc\,{\sc iii} & Ti\,{\sc iii} & Ti\,{\sc iv} & V\,{\sc iii} & Cr\,{\sc iii} & Fe\,{\sc iii} & Co\,{\sc iii} & Zn\,{\sc iii}\\
\noalign{\smallskip}
\hline
\noalign{\smallskip}
HE\,0151$-$3919          & $<5.10$      & $<5.80$       & $-$          & $7.50$       & $6.00$        & $<6.97$       & $<6.68$       & $<7.44$ \\
EC\,21494$-$7018         & $<6.42$      & $<7.78$       & $-$          & $7.50$       & $-$           & $7.79\pm0.30$ & $-$           & $-$   \\
EC\,15103$-$1557         & $<5.55$      & $<6.52$       & $-$          & $<7.28$      & $-$           & $7.64\pm0.23$ & $-$           & $-$   \\
EC\,11349$-$2753         & $<5.27$      & $6.15\pm0.64$ & $-$          & $<6.78$      & $-$           & $7.57\pm0.15$ & $-$           & $-$   \\
HD\,4539                 & $<5.29$      & $6.25\pm0.49$ & $<7.89$      & $<6.76$      & $-$           & $7.38\pm0.15$ & $-$           & $-$   \\
HE\,2307$-$0340          & $<5.49$      & $<6.42$       & $-$          & $7.40$       & $<5.84$       & $7.75\pm0.17$ & $<7.00$       & $<7.57$ \\
HE\,0539$-$4246          & $<5.50$      & $6.50$        & $-$          & $7.50$       & $6.45\pm0.49$ & $7.57\pm0.25$ & $<7.34$       & $<7.62$ \\
EC\,14345$-$1729         & $<5.52$      & $<7.34$       & $-$          & $<7.10$      & $-$           & $7.84\pm0.27$ & $-$           & $-$ \\
PG\,1627$+$017$^{\rm r,p}$ & $-$        & $<7.28$       & $-$          & $<7.02$      & $-$           & $7.70\pm0.10$ & $-$           & $-$ \\
PG\,1432$+$004           & $<5.28$      & $6.25\pm0.49$ & $-$          & $<6.85$      & $-$           & $7.14\pm0.16$ & $-$           & $-$ \\
HE\,2208$+$0126$^{\rm r}$          & $<5.44$      & $<6.36$       & $-$          & $7.20$       & $6.10\pm0.42$ & $7.30\pm0.28$ & $<7.30$       & $<7.10$ \\
GD\,687$^{\rm r}$        & $<5.65$      & $<6.50$       & $-$          & $7.40$       & $<6.00$       & $7.00$        & $<7.20$       & $<7.50$ \\
EC\,20106$-$5248         & $<5.57$      & $<7.35$       & $-$          & $<7.09$      & $-$           & $7.71\pm0.20$ & $-$           & $-$ \\
BD$+$48\,2721            & $<4.69$      & $<5.38$       & $-$          & $<6.82$      & $-$           & $7.00\pm0.35$ & $-$           & $-$ \\
HD\,205805               & $<4.80$      & $6.10\pm0.71$ & $<7.64$      & $<6.24$      & $-$           & $7.38\pm0.14$ & $-$           & $-$ \\
PB\,7352$^{\rm r}$       & $<5.28$      & $-$           & $-$          & $<6.99$      & $-$           & $7.65\pm0.13$ & $-$           & $-$ \\
TON\,S\,135$^{\rm r}$    & $<5.65$      & $<5.90$       & $-$          & $<7.17$      & $-$           & $7.81\pm0.17$ & $-$           & $-$ \\
HE\,0321$-$0918          & $5.40$       & $6.70$        & $-$          & $7.40$       & $6.60\pm0.42$ & $7.37\pm0.06$ & $<7.40$       & $<7.40$ \\
LB\,1516$^{\rm r,l}$     & $<5.91$      & $<6.40$       & $-$          & $7.20$       & $-$           & $7.76\pm0.24$ & $-$           & $-$ \\ 
PG\,0001$+$275           & $<5.30$      & $-$           & $-$          & $<7.03$      & $-$           & $7.70\pm0.24$ & $-$           & $-$ \\ 
PG\,1653$+$131$^{\rm r}$ & $-$          & $<6.39$       & $-$          & $<7.41$      & $-$           & $7.74\pm0.17$ & $-$           & $-$ \\
HE\,0532$-$4503$^{\rm r}$ & $<4.98$     & $<5.38$       & $-$          & $6.70$       & $5.50$        & $7.37\pm0.17$ & $<6.49$       & $-$ \\ 
PG\,0342$+$026           & $-$          & $6.10\pm0.57$ & $-$          & $<6.75$      & $-$           & $7.63\pm0.18$ & $-$           & $-$ \\
GD\,108                  & $<5.25$      & $<5.57$       & $-$          & $<6.91$      & $-$           & $7.21\pm0.12$ & $-$           & $-$ \\
Feige\,65                & $<5.33$      & $-$           & $-$          & $<7.14$      & $-$           & $7.74\pm0.21$ & $-$           & $-$ \\
JL\,82$^{\rm r,l}$         & $<5.38$      & $<6.59$       & $-$          & $<7.16$      & $-$           & $7.66\pm0.22$ & $-$           & $-$ \\
PHL\,457$^{\rm r}$       & $<5.46$      & $<5.62$       & $-$          & $<6.96$      & $-$           & $7.85\pm0.23$ & $-$           & $-$ \\
PG\,1248$+$164$^{\rm r}$ & $-$          & $<7.39$       & $-$          & $<7.20$      & $-$           & $7.45\pm0.49$ & $-$           & $-$ \\
PHL\,44$^{\rm l}$        & $<5.56$      & $<5.68$       & $-$          & $<7.06$      & $-$           & $7.65\pm0.22$ & $-$           & $-$ \\
PG\,1432$+$159$^{\rm r}$ & $<5.61$      & $<7.33$       & $-$          & $<7.06$      & $-$           & $7.25\pm0.07$ & $-$           & $-$ \\
SB\,815                  & $<5.56$      & $<5.69$       & $-$          & $<6.93$      & $-$           & $7.56\pm0.15$ & $-$           & $-$ \\
PG\,2205$+$023           & $<5.74$      & $6.30$        & $-$          & $<7.13$      & $-$           & $8.10$        & $-$           & $-$ \\
PG\,2314$+$076           & $<5.36$      & $<5.74$       & $-$          & $<7.07$      & $-$           & $7.03\pm0.31$ & $-$           & $-$ \\
PG\,1716$+$426$^{\rm r,l}$ &  $<5.63$   & $<7.33$       & $-$          & $<7.06$      & $-$           & $7.87\pm0.31$ & $-$           & $-$ \\
CPD$-$64$^{\rm \circ}$481$^{\rm r}$ & $<4.82$ & $5.95\pm0.64$ & $<7.46$ & $<6.22$     & $-$           & $7.41\pm0.17$ & $-$           & $-$ \\
PG\,0101$+$039$^{\rm r}$ & $<5.33$      & $<5.66$       & $-$          & $<6.99$      & $-$           & $7.55\pm0.22$ & $-$           & $-$ \\
PG\,2345$+$318$^{\rm r}$ & $-$          & $-$           & $-$          & $<6.97$      & $-$           & $7.17\pm0.06$ & $-$           & $-$ \\
PG\,1743$+$477$^{\rm r}$ & $-$          & $<6.71$       & $-$          & $<6.75$      & $-$           & $7.17\pm0.29$ & $-$           & $-$ \\
TON\,S\,183$^{\rm r}$    & $<5.31$      & $<5.63$       & $-$          & $<6.92$      & $-$           & $7.51\pm0.21$ & $-$           & $-$ \\ 
HS\,2357$+$2201$^{\rm r}$ &  $<5.41$    & $<5.83$       & $-$          & $<7.02$      & $<5.81$       & $7.33\pm0.28$ & $<6.99$       & $-$ \\
EC\,14338$-$1445$^{\rm r}$ & $<5.92$    & $6.55\pm0.78$ & $-$          & $<7.27$      & $-$           & $7.25\pm0.23$ & $-$           & $-$ \\ 
HD\,171858$^{\rm r}$     & $-$          & $<5.35$       & $<7.89$      & $<6.75$      & $-$           & $7.59\pm0.20$ & $-$           & $-$ \\
SB\,485                  & $<5.23$      & $<6.00$       & $-$          & $7.21$       & $<5.50$       & $7.38\pm0.15$ & $<6.55$       & $<6.40$ \\ 
EC\,03591$-$3232         & $<5.33$      & $6.80\pm0.42$ & $<7.67$      & $<6.77$      & $-$           & $7.03\pm0.19$ & $-$           & $-$ \\
\noalign{\smallskip}
\hline
\noalign{\smallskip}
Object             & Sc\,{\sc iii} & Ti\,{\sc iii} & Ti\,{\sc iv} & V\,{\sc iii} & Cr\,{\sc iii} & Fe\,{\sc iii} & Co\,{\sc iii} & Zn\,{\sc iii}\\
\noalign{\smallskip}
\hline
\noalign{\smallskip}
EC\,12234$-$2607         & $<6.05$      & $<6.44$       & $-$          & $7.60$       & $-$           & $7.10\pm0.21$ & $-$           & $-$ \\ 
PG\,2349$+$002           & $<5.70$      & $6.00$        & $<8.20$      & $<7.04$      & $-$           & $7.26\pm0.20$ & $-$           & $-$ \\ 
HE\,0136$-$2758          & $<5.45$      & $<5.98$       & $<5.98$      & $<6.94$      & $5.70$        & $7.38\pm0.17$ & $7.20\pm0.28$ & $<6.51$ \\
PG\,1549$-$001           & $<5.47$      & $7.10\pm0.30$ & $-$          & $7.40$       & $6.81$        & $7.07\pm0.15$ & $<6.90$       & $<6.50$ \\
HE\,0016$+$0044          & $5.70$       & $6.40$        & $-$          & $7.40$       & $6.05\pm0.49$ & $6.85\pm0.07$ & $<6.85$       & $<6.49$ \\
HE\,2349$-$3135          & $<5.71$      & $7.10\pm0.14$ & $<6.43$      & $<7.00$      & $<6.06$       & $6.93\pm0.21$ & $<7.44$       & $<6.72$ \\
PG\,1725$+$252$^{\rm r}$ & $-$          & $-$           & $<7.76$      & $<6.98$      & $-$           & $7.53\pm0.33$ & $-$           & $-$     \\
HE\,0007$-$2212          & $<5.53$      & $6.70\pm0.54$ & $<7.70$      & $6.80\pm0.14$ & $6.50\pm0.28$ & $7.17\pm0.15$ & $<7.31$      & $<6.20$ \\
PG\,1329$+$159$^{\rm r}$ & $-$          & $<7.29$       & $-$          & $<7.09$      & $-$           & $7.53\pm0.15$ & $-$           & $-$  \\
LB\,275                  & $6.20$       & $6.82\pm0.45$ & $<7.98$      & $<7.07$      & $<6.88$       & $7.03\pm0.10$ & $<7.28$       & $<6.30$ \\
Feige\,48$^{\rm r}$      & $-$          & $<5.74$       & $-$          & $<6.97$      & $-$           & $7.46\pm0.21$ & $-$           & $-$ \\
HE\,1421$-$1206$^{\rm r}$ & $<5.59$     & $6.63\pm0.47$ & $<6.59$      & $<7.17$      & $7.10\pm0.14$ & $7.37\pm0.15$ & $<6.90$       & $-$ \\
PG\,0133$+$114$^{\rm r}$ & $5.60$       & $6.00$        & $-$          & $<6.81$      & $-$           & $7.28\pm0.12$ & $-$           & $-$ \\
PG\,1101$+$249$^{\rm r}$ & $<5.39$      & $-$           & $<7.48$      & $<6.81$      & $-$           & $7.25\pm0.21$ & $-$           & $-$ \\
HE\,0929$-$0424$^{\rm r}$ & $5.40$  & $6.87\pm0.15$ & $7.65\pm0.21$ & $7.60\pm0.28$ & $6.30$      & $7.10$        & $<7.28$       & $6.20$ \\
PHL\,861$^{\rm r}$       & $<5.71$      & $6.80\pm0.14$ & $8.20\pm0.14$ & $<7.35$     & $7.35\pm0.35$ & $7.10\pm0.28$ & $<7.38$       & $<6.70$ \\
PG\,1303$+$097           & $<5.80$      & $<7.52$       & $-$          & $<7.56$      & $-$           & $7.23\pm0.31$ & $-$           & $-$ \\
HE\,2150$-$0238$^{\rm r}$ & $5.70$  & $7.03\pm0.46$ & $-$          & $<7.10$      & $6.95\pm0.07$ & $7.00$        & $<7.41$       & $<6.69$ \\
PG\,1512$+$244$^{\rm r}$ & $<5.37$      & $-$           & $<7.38$      & $-$          & $-$           & $<6.49$       & $-$           & $-$ \\
HE\,2135$-$3749$^{\rm r}$ & $5.00$  & $6.60\pm0.27$ & $<7.37$      & $6.80$       & $6.90\pm0.14$ & $6.65\pm0.35$ & $<6.53$       & $-$ \\
HE\,2222$-$3738          & $5.70$       & $6.77\pm0.47$ & $<7.49$      & $7.27$       & $6.85\pm0.07$ & $6.93\pm0.40$ & $<6.90$       & $<6.54$ \\
HE\,1047$-$0436$^{\rm r}$ & $5.90$  & $7.03\pm0.41$ & $7.00$       & $7.20$       & $7.10\pm0.42$ & $6.90\pm0.36$ & $<6.92$       & $-$ \\
PG\,1519$+$640$^{\rm r}$ & $<5.62$      & $<5.72$       & $-$          & $<7.03$      & $-$           & $<7.03$       & $-$           & $-$ \\
HE\,2238$-$1455          & $<5.59$      & $6.70$        & $<6.31$      & $<6.99$      & $6.70\pm0.28$ & $<6.70$       & $<6.95$       & $-$ \\
Feige\,38                & $-$          & $6.85\pm0.35$ & $<7.64$      & $<6.83$      & $-$           & $7.16\pm0.11$ & $-$           & $-$ \\
PG\,1710$+$490           & $-$          & $7.20\pm0.28$ & $7.70$       & $<7.30$      & $-$           & $7.75\pm0.07$ & $-$           & $-$ \\
EC\,14248$-$2647         & $<5.73$      & $6.75\pm0.64$ & $7.00$       & $<7.03$      & $-$           & $7.05\pm0.24$ & $-$           & $-$ \\
HE\,0230$-$4323$^{\rm r,s}$   & $<5.65$ & $6.95\pm0.07$ & $<7.46$      & $7.84$       & $6.70$        & $7.43\pm0.17$ & $<7.28$       & $<5.90$ \\ 
KPD\,2109$+$440$^{\rm s}$     & $-$     & $<6.32$       & $-$          & $<7.31$      & $-$           & $7.30\pm0.16$ & $-$           & $-$ \\ 
$[$CW83$]$\,1758$+$36         & $-$     & $6.95\pm0.49$ & $<7.61$      & $<7.47$      & $-$           & $7.17\pm0.06$ & $-$           & $-$ \\
EC\,21043$-$4017         & $<6.37$      & $<7.53$       & $-$          & $<7.45$      & $-$           & $<6.98$       & $-$           & $-$ \\
EC\,20229$-$3716         & $-$          & $<5.93$       & $<6.46$      & $7.10$       & $-$           & $7.21\pm0.24$ & $-$           & $-$ \\
PG\,1116$+$301$^{\rm r}$ & $-$          & $<7.54$       & $-$          & $<7.83$      & $-$           & $7.60\pm0.28$ & $-$           & $-$ \\
PG\,2151$+$100           & $<6.36$      & $-$           & $<7.11$      & $<7.56$      & $-$           & $<6.86$       & $-$           & $-$ \\
HS\,2033$+$0821          & $6.15\pm0.07$ & $7.50\pm0.46$ & $7.85\pm0.21$ & $<7.25$     & $7.00$        & $7.43\pm0.23$ & $<7.40$       & $-$ \\
EC\,05479$-$5818         & $6.30$       & $7.20\pm0.57$ & $7.15\pm0.07$ & $<7.30$     & $-$           & $<6.99$       &  $-$           & $-$ \\
PHL\,932                 & $<5.69$      & $7.30\pm0.14$ & $6.85\pm0.21$ & $<7.21$     & $7.10$        & $6.90\pm0.28$ &  $-$          & $6.00$ \\
EGB\,5$^{\rm r}$         & $-$          & $<6.22$       & $<6.22$       & $<7.09$     & $-$           & $<6.72$       &  $-$          & $-$ \\
PHL\,555                 & $<5.99$      & $7.10\pm0.57$ & $6.95\pm0.21$ & $<6.88$     & $7.35\pm0.49$ & $7.25\pm0.35$ &  $-$          & $<6.47$ \\
PG\,1219$+$534$^{\rm s}$ & $<6.29$      & $<6.37$       & $-$           & $<7.42$     & $-$           & $7.20$        &  $-$          & $-$ \\
HE\,1050$-$0630          & $<5.80$      & $6.85\pm0.35$ & $7.05\pm0.07$ & $7.60$      & $7.30$        & $7.30\pm0.28$ &  $-$          & $<6.50$ \\
HE\,1450$-$0957          & $<6.00$      & $<6.82$       & $<7.70$       & $<8.06$     & $<7.24$       & $7.10\pm0.28$ &  $-$          & $-$ \\
EC\,13047$-$3049         & $<6.36$      & $-$           & $<6.65$       & $<7.84$     & $-$           & $7.40$        &  $-$          & $-$ \\ 
HE\,1448$-$0510$^{\rm r}$  & $<5.80$    & $6.87\pm0.50$ & $6.90$        & $<7.88$     & $<6.50$       & $7.20$        &  $-$          & $-$ \\
PHL\,334                 & $-$          & $<6.37$       & $-$           & $<7.23$     & $-$           & $<6.82$       &  $-$          & $-$ \\
\noalign{\smallskip}
\hline
\noalign{\smallskip}
Object             & Sc\,{\sc iii} & Ti\,{\sc iii} & Ti\,{\sc iv} & V\,{\sc iii} & Cr\,{\sc iii} & Fe\,{\sc iii} & Co\,{\sc iii} & Zn\,{\sc iii}\\
\noalign{\smallskip}
\hline
\noalign{\smallskip}
HE\,2151$-$1001          & $<6.47$      & $<6.71$       & $<7.00$       & $<7.96$     & $7.20$        & $7.15\pm0.21$ &  $-$          & $-$ \\
Feige\,49                & $-$          & $-$           & $<6.47$       & $<7.78$     & $-$           & $<7.02$       &  $-$          & $-$ \\
PG\,0909$+$164           & $-$          & $<6.52$       & $-$           & $<7.50$     & $-$           & $<6.94$       &  $-$          & $-$ \\
CD$-$24$^{\rm \circ}$731$^{\rm r}$ & $-$ & $<6.48$      & $<6.30$       & $<8.10$     & $-$           & $<7.50$       &  $-$          & $-$ \\
HE\,1021$-$0255          & $<6.42$      & $-$           & $<6.67$       & $<7.89$     & $7.70$        & $7.80\pm0.01$ &  $-$          & $<6.71$ \\
HD\,149382               & $-$          & $<7.36$       & $6.80$        & $<7.19$     & $-$           & $<6.81$       &  $-$          & $-$ \\
PG\,0909$+$276           & $7.00$       & $8.00\pm0.42$ & $8.35\pm0.21$ & $<8.00$     & $-$           & $<7.87$       &  $-$          & $-$ \\
HE\,0101$-$2707          & $6.30\pm0.42$ & $7.55\pm0.07$ & $6.85\pm0.35$ & $<7.40$    & $7.10$        & $<7.34$       &  $-$          & $-$ \\
HE\,0019$-$5545          & $<6.50$      & $<6.79$       & $6.70\pm0.42$ & $7.90$      & $7.20$        & $7.20$        &  $-$          & $-$ \\
PG\,1207$-$032           & $<6.00$      & $7.69$        & $6.90\pm0.14$ & $7.55\pm0.07$ & $7.60$      & $7.40$        &  $-$          & $<6.70$ \\
GD\,619                  & $<6.00$      & $7.05\pm0.07$ & $6.65\pm0.21$ & $<7.70$     & $7.60$        & $7.59\pm0.10$ &  $-$          & $-$ \\
HE\,0123$-$3330          & $<6.10$      & $7.50\pm0.28$ & $-$           & $<7.92$     & $7.40$        & $7.30\pm0.14$ &  $-$          & $-$ \\
PG\,1505$+$074           & $-$          & $-$           & $<5.90$       & $<7.09$     & $-$           & $<6.72$       &  $-$          & $-$ \\
HE\,1407$+$0033          & $<6.51$      & $7.15\pm0.35$ & $<6.63$       & $7.90$      & $<7.54$       & $<7.29$       &  $-$          & $<6.70$ \\
PHL\,1548                & $-$          & $6.70$        & $6.95\pm0.07$ & $<7.97$     & $-$           & $<7.87$       &  $-$          & $-$ \\
$[$CW83$]$\,0512$-$08    & $-$          & $7.50$        & $7.70\pm0.14$ & $<7.94$     & $-$           & $<7.81$       &  $-$          & $-$ \\ 
PB\,5333                 & $-$          & $-$           & $<6.55$       & $<7.92$     & $-$           & $<7.32$       &  $-$          & $-$ \\ 
\noalign{\smallskip}
\hline
\end{longtable}
\tablefoot{$^{\rm r}$Radial velocity variable sdB. $^{\rm s}$Short-period pulsator of sdBV$_{\rm r}$-type. $^{\rm l}$Long-period pulsator of sdBV$_{\rm s}$-type.}

\newpage

\end{landscape}

\end{appendix}
\end{document}